# Designing magnetic properties in CrSBr through hydrostatic pressure and ligand substitution

*Evan J. Telford[Δ], Daniel G. Chica[Δ], Kaichen Xie, Nicholas S. Manganaro, Chun-Ying Huang, Jordan Cox, Avalon H. Dismukes, Xiaoyang Zhu, James P. S. Walsh, Ting Cao, Cory R. Dean, Xavier Roy*, Michael E. Ziebel**

Evan J. Telford, Daniel G. Chica, Chun-Ying Huang, Jordan Cox, Avalon H. Dismukes, Prof. Xiaoyang Zhu, Prof. Xavier Roy, Michael E. Ziebel
Department of Chemistry, Columbia University, New York, NY, USA
E-mail : xr2114@columbia.edu, mez2127@columbia.edu

Evan J. Telford, Prof. Cory R. Dean
Department of Physics, Columbia University, New York, NY, USA

Kaichen Xie, Prof. Ting Cao
Department of Materials Science and Engineering, University of Washington, Seattle, WA, USA

Nicholas S. Manganaro, Prof. James P. S. Walsh
Department of Chemistry, University of Massachusetts Amherst, Amherst, MA, USA

Δ These authors contributed equally to this work

**Abstract:**
The ability to control magnetic properties of materials is crucial for fundamental research and underpins many information technologies. In this context, two-dimensional materials are a particularly exciting platform due to their high degree of tunability and ease of implementation into nanoscale devices. Here we report two approaches for manipulating the A-type antiferromagnetic properties of the layered semiconductor CrSBr through hydrostatic pressure and ligand substitution. Hydrostatic pressure compresses the unit cell, increasing the interlayer exchange energy while lowering the Néel temperature. Ligand substitution, realized synthetically through Cl alloying, anisotropically compresses the unit cell and suppresses the Cr-halogen covalency, reducing the magnetocrystalline anisotropy energy and decreasing the Néel temperature. A detailed structural analysis combined with first-principles calculations reveal that alterations in the magnetic properties are intricately related to changes in direct Cr–Cr exchange interactions and the Cr–anion superexchange pathways. Further, we demonstrate that Cl alloying enables chemical tuning of the interlayer coupling from antiferromagnetic to ferromagnetic, which is unique amongst known two-dimensional magnets. The magnetic tunability, combined with a high ordering temperature, chemical stability, and functional semiconducting properties, make CrSBr an ideal candidate for pre- and post-synthetic design of magnetism in two-dimensional materials.

**Main Text:**

The discovery of two-dimensional (2D) magnets[1, 2], prepared by mechanical exfoliation of bulk van der Waals (vdW) materials, provides an ideal platform to understand and ultimately control 2D magnetism, fueling opportunities for atomically-thin spintronic[3] and magneto-optic devices[4, 5]. Among the growing number of 2D vdW magnets, including binary metal halides[6-9] and chalcogenides[10, 11], MXenes[12], and transition metal ternary compounds[13-17], semiconductors have generated intense interest as they offer the possibility of strongly coupling magnetic order to both optical and electronic properties. The ternary vdW A-type antiferromagnet CrSBr is a particularly exciting material boasting a high Néel temperature $T_N$ = 132 K, stability under ambient conditions[17, 18], and functional semiconducting transport properties[18-22]. Furthermore, CrSBr manifests a uniquely strong coupling between magnetism and electronic[18, 19], optical[17, 23], and structural properties[24], as well as tunable coupling between magnons and excitons[25, 26]. Consequently, developing routes to modify the magnetic properties of CrSBr could unlock new magneto-optical, magneto-electric, magneto-elastic and quantum transduction phenomena.

In the insulating $CrX_3$ (X = Cl, Br, I) family of vdW magnets, previous efforts have demonstrated the ability to tune magnetic properties both at the few-layer limit and in the bulk, via electrostatic doping[27], electric-field tuning[28], hydrostatic pressure[29-31], strain[32], and halogen substitution[33, 34]. Accessing similarly tunable magnetism in CrSBr, while also leveraging its enhanced ordering temperature, chemical stability, semiconducting properties, and strong coupling between magnetic order and other physical properties, could greatly improve the viability of vdW magnets for use in nanoscale spintronics. Recent experiments demonstrated that uniaxial strain on thin flakes of CrSBr changed the magnetic ground state from antiferromagnetic (AFM) to ferromagnetic (FM) due to a change in the sign of the interlayer coupling[24]. However, there have been no experimental investigations of strategies to tune the *intralayer* coupling in CrSBr, which is expected to most strongly affect its magnetic ordering temperature. Understanding how structural and electronic modifications to CrSBr affect these intralayer magnetic properties will enable the engineering of new materials in this class of transition-metal ternary compounds with designer magneto-electronic and magneto-optical properties.

In this work, we uncover how physical and chemical modifications of the CrSBr structure affect the magnetic properties through combined magnetic, structural, and computational analysis. We find that compression of the lattice under hydrostatic pressure (**figure 1A**) reduces $T_N$ through suppression of intralayer FM interactions. Upon Cl alloying, the combined effects of anisotropic lattice compression (**figure 1B**) and reduced Cr–halogen covalency lead to an even larger decrease in $T_N$. In both cases, the reduced ordering temperature comes from suppressed intralayer FM superexchange interactions, highlighting the delicate balance between Cr–Cr direct exchange and Cr–anion superexchange pathways. At the highest accessible Cl content, we observe suppressed in-plane magnetic anisotropy and a glassy magnetic ground state, hosting competing FM and AFM interlayer interactions, which could provide a platform for exploring phase transitions between FM and AFM states with external stimuli. Together, these results reveal a rich magnetic phase space in the CrSBr family, motivating further implementation of these materials into 2D spintronic devices.

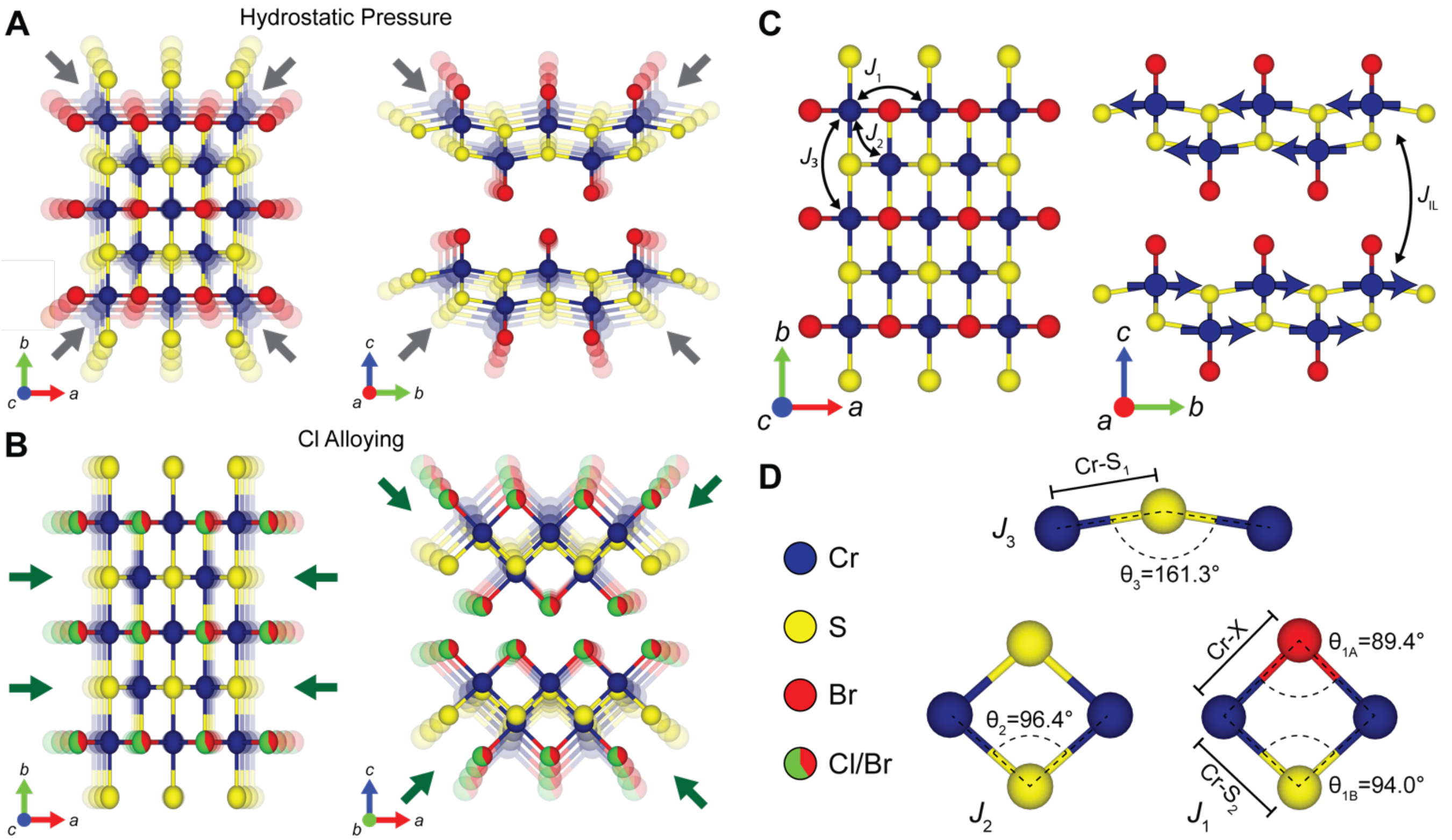

**Figure 1: Structure and magnetic coupling of CrSBr under hydrostatic pressure and Cl alloying.** **A)** Schematic depicting the effect of hydrostatic pressure on the crystal structure of CrSBr as viewed along the *c*-axis (left) and *a*-axis (right). The dark gray arrows denote the qualitative direction of the unit-cell changes. **B)** Schematic depicting the effect of Cl alloying on the crystal structure of $CrSBr_{1-x}Cl_x$ as viewed along the *c*-axis (left) and *b*-axis (right). Dark green arrows denote the qualitative direction of the unit-cell changes. **C)** Crystal structure of CrSBr as viewed along the *c*-axis (left) and *a*-axis (right). Double ended black arrows denote the direction of the relevant magnetic couplings. Dark blue arrows denote the direction of $Cr^{3+}$ moments in the antiferromagnetic state. **D)** Schematic of the superexchange pathways for the three largest intraplanar FM couplings in CrSBr. The corresponding bonds and bond angles contributing to the superexchange interactions are labeled.

The structure of a vdW CrSBr layer consists of two buckled rectangular planes of CrS fused together, with both surfaces capped by Br atoms (**figure 1C**). Stacking of the layers along the *c*-axis produces an orthorhombic structure with the space group *Pmmn*[18, 35, 36]. The primary magnetic couplings consist of three intralayer FM superexchange interactions (denoted $J_1$, $J_2$, and $J_3$) mediated by intralayer Cr–S–Cr and Cr–Br–Cr bonds[37-43] (**figure 1D**). The interlayer AFM super-superexchange coupling ($J_{IL}$) is mediated by Cr–Br–Br–Cr interactions between the sheets (**figure 1C**)[17, 23, 43, 44]. The strong intralayer coupling gives rise to short-range FM correlations below a characteristic temperature ($T_c$ ~ 160 K)[17, 18, 42, 43], while the weaker interlayer exchange (**table s1**) induces long-range A-type AFM order below $T_N$ = 132 K[43]. In the magnetically ordered state, each layer orders ferromagnetically with adjacent layers aligned antiferromagnetically along the stacking direction (**figure 1C**)[18, 35, 42, 43]. CrSBr exhibits uniaxial magnetic anisotropy along the *b*-axis, originating from anisotropic exchange interactions mediated by the surface-capping Br[42, 45, 46].

While intralayer superexchange interactions in CrSBr are FM, analogous interactions in the isostructural compounds VOCl[47], CrOCl[48], and FeOCl[49] are AFM, suggesting that the sign of magnetic exchange in this family of materials may be highly sensitive to Cr–halogen–Cr and Cr–chalcogen–Cr bond angles. With this in mind, we chose hydrostatic pressure as an initial route to modify the magnetic properties of CrSBr, as pressure provides a medium to modify the structure without changing chemical properties. For measurements of CrSBr under hydrostatic pressure (*P*), samples were prepared by grinding bulk single crystals in liquid nitrogen (see methods for details). The powder was then mixed with Daphne oil and loaded into a high-pressure cell along with a small piece of Pb acting as a manometer (**figures s1-s3** and methods for details). We performed magnetic measurements on the randomly-oriented powder as a function of temperature (*T*), magnetic field ($\mu_0 H$), and *P*.

**Figure 2A** presents the magnetic susceptibility ($\chi$) of CrSBr versus *T* for various *P* up to 1.39 GPa. $T_N$ manifests as a peak in $\chi$ vs. *T* and is extracted numerically by finding the zero-crossings of $d\chi/dT$ (**figure s4**). The ambient-*P* $T_N$ = 135 ± 3 K is in agreement with previous reports[17-20, 23, 35, 42, 43, 50-52]. Upon the application of *P*, $T_N$ decreases linearly at a rate of $dT_N/dP = -12.6 \pm 1.0$ K·GPa$^{-1}$ (inset of **figure 2A**). Curie-Weiss analysis reveals that the Weiss temperature ($\theta_W$) also decreases with increasing *P*, while the Curie constant (*C*) is independent of pressure (**figure s5**), indicating a weakening of the intralayer FM coupling strength with no change in the $S = {}^3/_2$ $Cr^{3+}$ moments. Further measurements of $\chi$ vs *T* with a large applied $\mu_0 H$ = 3 T (where all spins in the magnetic state are polarized along the field direction) (**figure s6**) show a paramagnetic-to-FM phase transition with a decreasing Curie temperature with increasing *P*, supporting the conclusion that increasing *P* weakens the intralayer FM coupling. In **figure 2B**, we plot magnetization (*M*) versus $\mu_0 H$ with increasing *P*. Because the CrSBr samples were measured as a randomly-oriented powder, we expect the *M* vs. $\mu_0 H$ traces to be an average of the axial-oriented *M* vs. $\mu_0 H$ traces (**figure s7, s8**). At ambient *P*, *M* vs. $\mu_0 H$ is approximately linear at low $\mu_0 H$ followed by a change in slope at $\mu_0 H$ = 0.28 ± 0.05 T (*b*-axis saturation field) and a subtle kink at $\mu_0 H$ = 0.46 ± 0.05 T (*a*-axis saturation field) followed by saturation at $\mu_0 H$ = 1.05 ± 0.05 T (*c*-axis saturation field). With increasing *P*, the low-field slope decreases, resulting in an increasing saturation magnetic field $H_{SAT}$ (defined here as the $\mu_0 H$ at which $M = 0.9 \cdot M_{SAT}$). $H_{SAT}$ increases at a rate of $dH_{SAT}/dP = 0.49 \pm 0.03$ T·GPa$^{-1}$ (inset of **figure 2B**). For consistency, we repeated all measurements on a second CrSBr growth batch, which show quantitatively similar results (insets of **figures 2A, B** and **figure s9**). We note that powder X-ray diffraction (PXRD) measurements showed no evidence of irreversible phase transitions after grinding or applying maximum $P$ (**figure s10**).

To interpret the changes in magnetic properties of CrSBr under $P$, we measured the lattice parameters of CrSBr under $P$ (**figures s11-s15** and methods for details) and performed complementary theoretical simulations to calculate both the relaxed structural parameters and magnetic properties of CrSBr under $P$ (see methods for details). In **figure 2C**, the experimental lattice parameters are plotted versus $P$. All lattice parameters decrease, with the most significant change along the *c*-axis. Our density functional theory calculations well-predict the experimental change in lattice parameters under $P$ (**figure s12**) and find that, as the *c*-axis compresses, the corresponding interlayer AFM coupling drastically strengthens (by 340% at 1.5 GPa - inset of **figure 2D**). From this, one might expect $T_N$ to *increase* under $P$. However, all primary intralayer FM couplings weaken ($J_1$, $J_2$, and $J_3$ - **figure 2D** and **table s1**) and the magnitude of the strongest intralayer coupling, $J_2$, is more than 30 times that of $J_{IL}$ for the entire $P$ range, indicating that the *intralayer* magnetic exchange is the dominant contribution to the ordering temperature. The calculations fully support this conclusion, correctly predicting a decreasing $T_N$ with increasing $P$ (**figure 2E** and **table s1**). Furthermore, the experimental observation of an increase in $H_{SAT}$ with increasing $P$ is explained by the strengthening of the interlayer AFM coupling, in agreement with our calculations (**figure 2E** and **table s1, figure s7**).

Using the computed high-pressure structures, we can begin to rationalize the observed magnetic properties and derive magneto-structural correlations for CrSBr. Looking first at the interlayer spacing, we find the calculated vdW gap decreases significantly (~10% at 1.5 GPa) with $P$, leading to an increase in Cr–Br–Br–Cr overlap and thus $J_{IL}$ (**table s1**). The intralayer magnetic exchange is more complex. The intralayer exchange interactions in CrSBr represent a competition between FM superexchange interactions and weaker AFM direct exchange interactions. Changes in the superexchange interactions should be explained by the Goodenough-Kanamori-Anderson rules[53-55] for a $Cr^{3+}$ ion. These would predict the strongest FM coupling for bond angles near 90° and the strongest AFM coupling for bond angles near 180°. In contrast, the strength of AFM direct exchange interactions increases exponentially as the distance between magnetic ions shrinks.

To understand the magnetic behavior of CrSBr under pressure, the effects of both direct exchange and superexchange must be considered. With increasing pressure, the magnitude of direct exchange should increase for $J_1$, $J_2$, and $J_3$, as all Cr–Cr distances ($d_{Cr–Cr}$) shrink (**table s1**). These changes should be most pronounced for $J_1$ and $J_2$, which have experimentally-determined $d_{Cr–Cr}$ of ~3.51 and ~3.59 Å, respectively, whereas $d_{Cr–Cr}$ for $J_3$ is much larger (~4.76 Å). Because $d_{Cr–Cr}$ remains well outside the range of Cr–Cr bonding for all pressures studied here, we would expect the direct exchange interactions to remain small relative to superexchange interactions, which agrees with our experimental and computational data where the net intralayer coupling remains FM. However, the relative changes in the calculated exchange energies at 1.5 GPa compared to ambient pressure ($\Delta J_1 \sim \Delta J_3 > \Delta J_2$) are inconsistent with the expectations for direct exchange alone ($\Delta J_1 \sim \Delta J_2 > \Delta J_3$), suggesting that superexchange pathways are also affected by lattice compression.

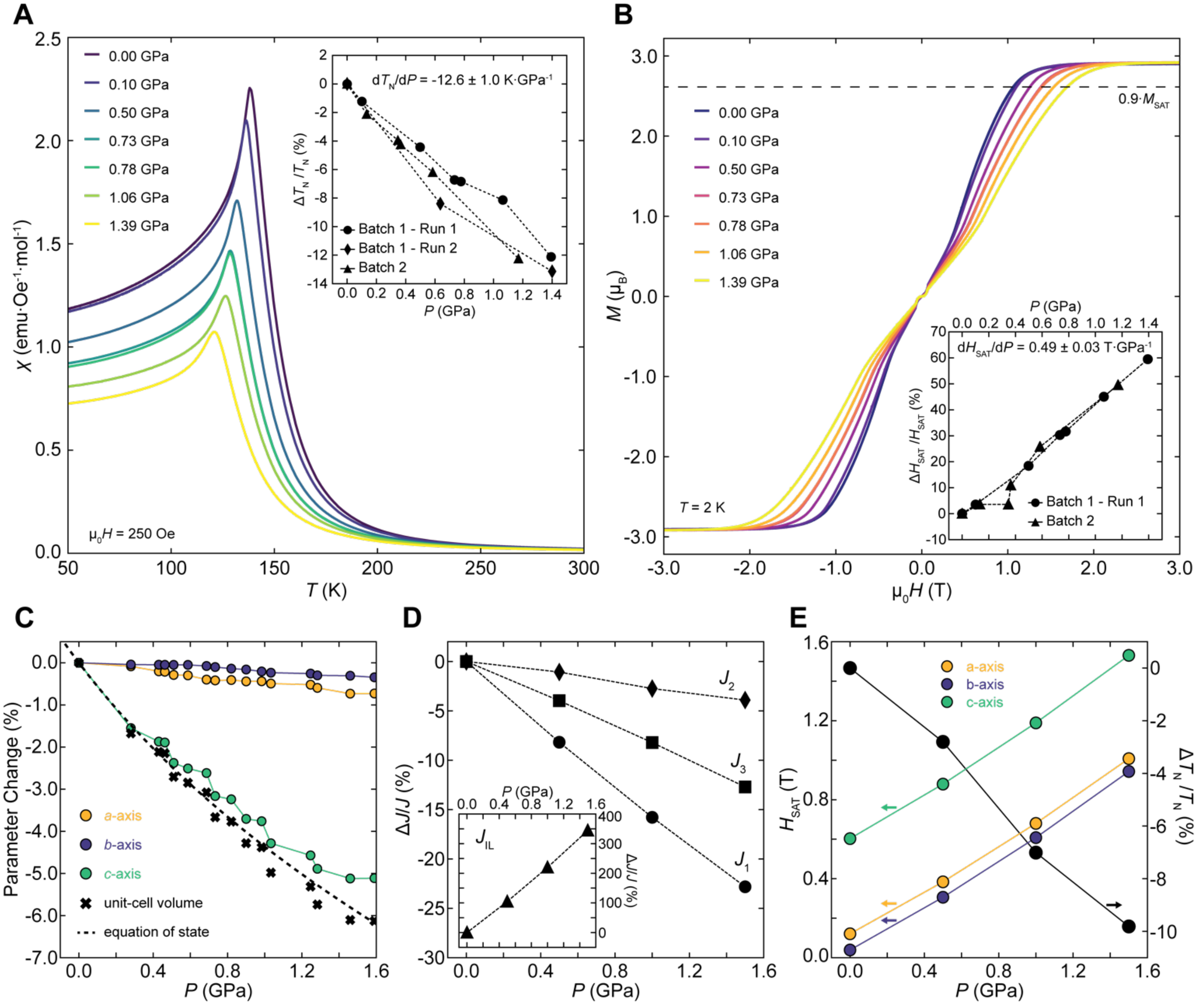

**Figure 2: Magnetic properties of CrSBr under hydrostatic pressure. A)** Zero-field-cooled magnetic susceptibility ($\chi$) versus temperature ($T$) for various applied hydrostatic pressures ($P$). A measuring field of 250 Oe was used for all traces. Inset shows extracted percentage change in $T_N$ versus $P$ for multiple measurement runs and growth batches. The extracted slope of $T_N$ versus $P$ is given in the inset. **B)** Magnetization ($M$) versus applied magnetic field ($\mu_0 H$) at 2 K for various $P$. $\mu_0 H$ is randomly oriented along all crystal axes. The saturation field ($H_{SAT}$) is defined as the $\mu_0 H$ at which $M$ is 90% of the saturation $M$ (denoted by a black dashed line). Inset shows extracted $H_{SAT}$ versus $P$ along with the extracted slope of $H_{SAT}$ versus $P$. **C)** Percentage change in lattice constants and the unit-cell volume versus $P$, as determined by powder X-ray diffraction. The dashed black line is a fit to an equation of state (see methods for details). **D)** Calculated percentage change in intralayer magnetic couplings versus $P$. Inset: calculated percentage change in interlayer magnetic coupling ($J_{IL}$) versus $P$. **E)** Calculated $H_{SAT}$ (left axis, orange, purple, and green dots) and $T_N$ (right axis, solid black dots) versus $P$.

As noted above, changes in superexchange pathways under pressure should be most sensitive to changes in the Cr–S–Cr and Cr–Br–Cr bond angles. At 1.5 GPa, all of these angles are predicted to change by less than 1° compared to the relaxed ambient-pressure structure, suggesting that the modulation of the superexchange energies should be smaller or similar in magnitude to the changes in direct exchange (**table s1**). The largest change is observed in the Cr–$S_1$–Cr bond angle ($\theta_3$, **figure 1D**), which increases towards 180°, enhancing the contribution of AFM exchange pathways and weakening the overall FM coupling (**table s1**). Consequently, both direct exchange and superexchange contributions contribute to the reduced magnitude of $J_3$ with increasing $P$. In contrast, for $J_1$ and $J_2$, all of the relevant Cr–S–Cr ($\theta_2$ and $\theta_{1B}$) and Cr–Br–Cr ($\theta_{1A}$) (**figure 1D**) bond angles trend towards 90° with increasing $P$ (**table s1**), which should enhance the FM superexchange interactions. Since the changes in bond angles are relatively small, the magnitude of these effects is likely minimized and could be less than the corresponding increase in AFM direct Cr–Cr exchange. Collectively, these results reveal the balance between superexchange and direct exchange that must be considered when designing new materials in this family.

While these results motivate studies of the magnetic behavior of CrSBr at even higher pressures where larger bond angle changes may affect superexchange pathways more drastically, chemical modification could induce larger structural changes than were obtained in the pressure range studied here. Specifically, we hypothesized that substitution of Br with Cl could induce a large lattice compression, while simultaneously allowing us to study the effects of changing Cr–halogen covalency on the magnetic properties. Furthermore, theoretical studies on ligand engineering[56] and strain[44] on chromium chalcohalides demonstrate changes to the magnetic properties with these perturbations. To explore this hypothesis, we synthesized a series of mixed-halogen compounds $CrSBr_{1-x}Cl_x$ with $x$ = 0 - 0.67 (from now on referred to as "Cl-$x$") using the chemical vapor transport approach (see methods for details). The crystal structure of each compound was determined through single-crystal X-ray diffraction (SCXRD) (**figure 3A** and **table s2**). Within the examined compositional range, the mixed-halogen alloys are isostructural to the parent compound CrSBr with the space group *Pmmn* (**figure 3A**). Because Cl is smaller than Br, Cl alloying has a significant impact on the lattice parameters, causing the lattice to "accordionize" along the *a*-axis, resulting in a decrease of the *a*- and *c*-lattice parameters with no significant change to the *b*-axis (**figure 3B** and **figure s16**). The incompressibility of the structure along the *b*-axis stems from the Cr–(Cl/Br) bonds lying parallel to the *ac*-plane. At the highest Cl content (Cl-67), the *a*- and *c*-axes have compressed by 2.2% and 4.9%, respectively, compared to CrSBr, with the *a*-axis compression exceeding the effects of pressure at 1.5 GPa (**figure s12**). We note that despite the structural changes resulting from Cl alloying, the crystals with the highest concentration of Cl remain exfoliatable down to the monolayer limit (**figure s17**).

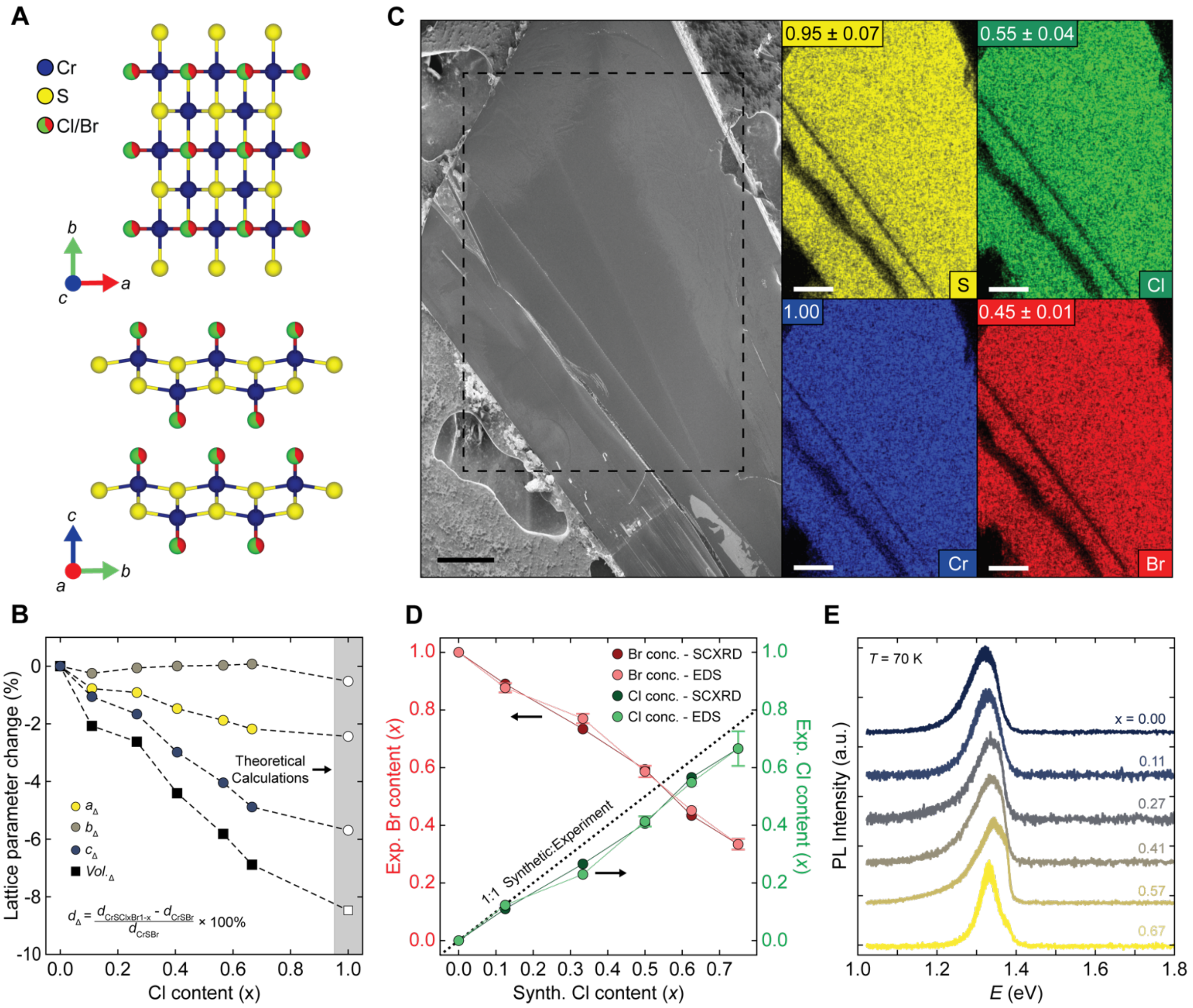

**Figure 3: Structural parameters and electronic properties of $CrSBr_{1-x}Cl_x$. A)** Crystal structure of Cl-57 as viewed along the *c*-axis (top) and *a*-axis (bottom). **B)** Lattice parameter ratio versus Cl content ($x$) for $CrSBr_{1-x}Cl_x$. **C)** Left: scanning electron microscopy (SEM) image of a cleaved crystal of Cl-57. Right: corresponding EDX elemental mapping. Blue, yellow, red, and green maps correspond to Cr, S, Br, and Cl elemental mapping, respectively. In each elemental map, the top-left inset shows the average concentration relative to Cr. The error bar is the standard deviation between multiple measurements and crystals. In all images, the scale bar is 100 µm. **D)** Halogen content determined using SCXRD and EDX versus Cl content used in chemical vapor transport reactions. The dashed black line demarcates 1:1 measured Cl content to Cl content used in chemical vapor transport reactions. **E)** Photoluminescence intensity versus photon energy for all synthesized Cl concentrations. The corresponding Cl content for each trace is given in the inset. All data were taken at 70 K.

The chemical compositions of all new materials were determined through a combination of refining the Cl/Br occupancy on the mixed anion site on SCXRD data and energy dispersive X-ray spectroscopy (EDX) (**figure 3C**, **figures s18-s23** and **table s3**). The percentages of Cl atoms substituted on the Br sites are close to the nominal stoichiometric amount of bromine and chlorine used in the synthesis (**figure 3D**). Importantly, the chemical composition maps measured using EDX (**figure 3C** and **figures s18-s23**) show no evidence of Cl or Br clustering on the micron scale. Polarized Raman spectroscopy on all alloys supports this, demonstrating a continuous frequency increase of characteristic CrSBr modes with increasing Cl concentration (**figure s24**), consistent with the homogeneous substitution of the lighter Cl atoms on Br sites[57-60]. Despite the significant structural changes upon Cl alloying, photoluminescence measurements on the various compositions show negligible changes in the optical band gap (**figure 3E**). This is consistent with previous band-structure calculations for CrSBr and CrSCl monolayers[61] and establishes our ability to tune the lattice and (as will be seen below) magnetic structure without significantly changing the electronic structure. Given the strong coupling between magnetism and optical and electronic properties in CrSBr, Cl alloying offers an entirely new space for designing magneto-optical and magneto-electronic properties without drastically affecting the band structure.

We now turn to explore how the magnetic properties of the mixed-halogen compounds change with increasing Cl content. In **figure 4A**, we plot $\chi$ vs $T$ for all compounds. For Cl concentrations up to Cl-41, we observe a clear AFM transition with a peak in $\chi$ at $T_N$, followed by a decrease in $\chi$ at low $T$ with no difference between zero-field-cooled (ZFC) and field-cooled (FC) traces. $T_N$ for each stoichiometry up to Cl-41 was extracted numerically by finding the zero-crossings in $d\chi/dT$ (**figure s25, s26**) and is found to decrease linearly at a rate of $dT_N/dx = -61.8$ $K \cdot x^{-1}$ (inset of **figure 4A**). The corresponding Curie-Weiss analysis (**Figure s27**) for this compositional range reveals that $\theta_W$ also decreases with increasing Cl content while the Curie constant remains constant, indicating a weakening of the intralayer FM coupling without a change in the $S = {}^3/_2$ $Cr^{3+}$ moments.

At high temperature, Cl-57 and Cl-67 follow a similar trend to the lower Cl concentrations. Specifically, $\theta_W$ lowers with increasing Cl content. Near the magnetic ordering temperature, however, the $\chi$ of Cl-57 and Cl-67 show distinctly different behavior from the lower Cl concentrations. For both compounds, the $\chi$ vs. $T$ traces display a small kink (at $T$ = 100 and 86 K for Cl-57 and Cl-67, respectively), a broad maximum (at $T$ = 89 K and 42 K), and a clear divergence between the FC and ZFC traces at low temperature. These features suggest the possibility of multiple magnetic phase transitions, and further indicate that the magnetic ground state of Cl-57 and Cl-67 cannot be described as a trivial antiferromagnet (**figure s26** for additional axial orientations). Complementary ac magnetic susceptibility measurements on Cl-57 and Cl-67 at zero dc field confirm the presence of multiple magnetic transitions and reveal frequency-dependent behavior (**figure s28, s29**), suggesting these compounds are best described as spin glasses or glassy magnets, wherein magnetic disorder emerges from competing FM and AFM interlayer interactions (see discussion below). Regardless, the magnetic critical temperatures (identified by peaks in the in-phase magnetic susceptibility) follow the same trend as the lower Cl concentrations (**figures s27-s29** for details). This indicates that, over the entire compositional range, increased Cl alloying leads to decreased magnetic ordering temperatures and weakened intralayer coupling.

To better understand the origin of this unusual magnetic behavior at high Cl content, we performed axial-oriented $M$ vs $\mu_0 H$ traces at 2 K for each stoichiometry (**figure 4B-D**). For $\mu_0 H$ along the easy *b*-axis (**figure 4C**), we observe a clear AFM-to-FM spin-flip transition for Cl doping up to Cl-41. The $H_{SAT}$, which we define as the midpoint of the transition where $M = 0.5 \cdot M_{SAT}$ to better illustrate the transition at higher Cl concentrations, decreases sharply with increasing Cl content, indicating a weakening of the interlayer AFM coupling. For Cl-57 and Cl-67, we observe s-shaped $M$ vs $\mu_0 H$ traces with no observable hysteresis. We propose that this change in behavior arises from competing interlayer FM Cr–Cl–Cl–Cr interactions and AFM Cr–Br–Br–Cr interactions. In the aggregate, this leads to negligible interlayer coupling for Cl-57 and Cl-67, and causes these two compositions to behave as ferromagnets under small applied fields. For $\mu_0 H$ along the *a*- and *c*-axes (**figure 4B** and **4D**, respectively), all alloys display similar behavior – a continuous spin canting process whereby the *b*-axis aligned spins cant towards the applied field direction. We observe a reduction in *a*- and *c*-axis $H_{SAT}$, defined as the point where $M = 0.9 \cdot M_{SAT}$, signifying a lowering of the magnetic anisotropy energy. A summary of the dependence of all axial saturation fields on Cl doping is given in the bottom inset of **figure 4A**. Remarkably, the *a*- and *b*-axis $H_{SAT}$ approach zero, indicating a diminishing anisotropy between the two in-plane directions, while the out-of-plane anisotropy only decreases by ~50% (see also **figure s30** for a detailed comparison between CrSBr and Cl-57). This reduction in the effective anisotropy between the *a*- and *b*-axes motivates further study of the critical behavior of these high Cl-content materials, specifically the possibility that they could display 2D-XY behavior at the monolayer limit[8, 33].

The large unit cells needed to adequately model random distributions of halogens in the alloys precluded detailed computational studies of specific compositions. Instead, we modeled the magnetic properties of the theoretical end-member of this series, CrSCl, to better understand the experimental trends. Because CrSCl is not currently experimentally accessible, we simulated and relaxed the structure using CrSBr as a model lattice (**figure s31** and **table s4**). The relaxed CrSCl structure agrees remarkably well with an extrapolation of the experimental data up to 100% Cl content (**figure 3B** and **table s4)**.

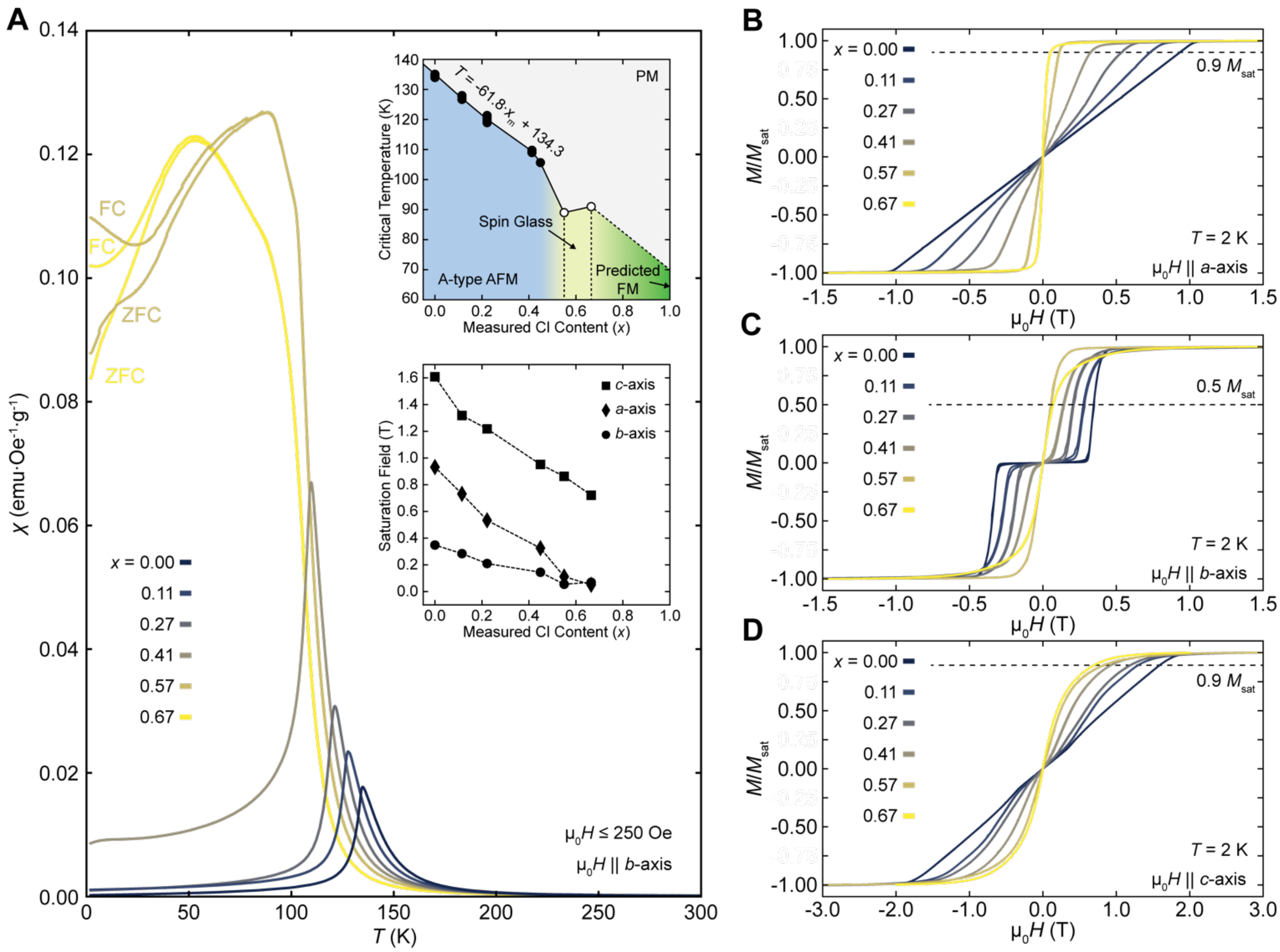

**Figure 4: Magnetic properties of $CrSBr_{1-x}Cl_x$. A)** Magnetic susceptibility (*χ)* versus temperature (*T*) for various Cl contents (*x*). A measuring field of 250 Oe was used for Cl-00, Cl-11, and Cl-27, whereas a measuring field of 100 Oe was used for Cl-41, Cl-57, and Cl-67. For Cl-00, Cl-11, Cl-27, and Cl-41, only the zero-field trace is shown as it overlaps the field-cooled trace. For Cl-57 and Cl-67, both zero-field-cooled and field-cooled traces are shown. Top inset shows extracted critical temperature versus Cl content. Grey, blue, and yellow regions correspond to experimentally identified PM, AFM, and spin-glass regions, respectively. Green region corresponds to the predicted FM state for CrSCl (**table s4**). Up to Cl-41 the critical temperature depends linearly on Cl doping. The linear fit parameters are given in the inset. Bottom inset shows the extracted saturation magnetic fields at 2 K for fields parallel to the *a*-, *b*-, and *c*-axes. **B-D)** Magnetization normalized to the saturation magnetization ($M/M_{SAT}$) versus applied magnetic field ($\mu_0 H$) at 2 K for $\mu_0 H$ oriented along the *a*- (**B**), *b*- (**C**), and *c*- (**D**) axes. The saturation magnetic field is defined as the magnetic field when the magnetization is 90%, 50%, and 90% of the saturation magnetization for the *a*-, *b*-, and *c*-axes, respectively.

As with the high-pressure data above, the combination of experimental magnetic data and computed magnetic and structural parameters allows us to derive magneto-structural correlations for halogen alloying in CrSBr. Increasing Cl content leads to a reduction in the interlayer spacing (**figure 3B)**, which could naively be expected to strengthen the interlayer magnetic exchange. Our experimental data, however, reveal that the interlayer coupling weakens with increasing Cl content (**figure 4C**). This behavior can be explained by the reduced orbital overlap of interlayer Cr–Cl–Cl–Cr exchange compared to Cr–Br–Br–Cr. Consistent with this hypothesis, calculations predict a change in the interlayer coupling from AFM in CrSBr to FM in CrSCl (**table s4** and **figure s31**), confirming that orbital overlap between the halogens across the vdW gap, rather than the interlayer Cr–Cr distance is responsible for directing the sign and strength of interlayer exchange. These results can also explain the glassy behavior of Cl-57 and Cl-67, which arises from competing interlayer FM Cr–Cl–Cl–Cr and AFM Cr–Br–Br–Cr interactions. We note that the change in sign of the interlayer coupling upon Cl substitution in CrSBr is distinctly different from what is observed in bulk chromium trihalides, where interlayer coupling is always AFM in the high temperature monoclinic structure and FM in the low temperature rhombohedral structure, independent of the identity of the halide[33]. The weak interlayer coupling emerging from competing FM and AFM interactions in Cl-57 and Cl-67 should make the magnetic ground state in these materials particularly susceptible to external stimuli, such as strain, pressure, and magnetic field, making these materials promising candidates for switchable 2D devices.

The largest change in the calculated intralayer coupling upon Cl substitution is the magnitude of $J_1$, which decreases by ~80% while remaining FM (**table s4**). The shorter calculated $d_{Cr\text{–}Cr}$ in CrSCl compared to CrSBr should increase the contributions of AFM direct exchange, though this is unlikely to fully explain the marked drop in the magnitude of the exchange energy. While the Cr–$S_2$–Cr and Cr–X–Cr bond angles associated with $J_1$ do change with halogen substitution (**figure s16** and **table s4**), the large reduction in the superexchange contribution to $J_1$ is most likely driven by the more ionic nature of the Cr–Cl bond compared to the more covalent Cr–Br bond. This predicted decrease in $J_1$ for CrSCl compared to CrSBr explains a majority of the reduction in $\theta_W$ with increasing Cl content. However, examination of the other exchange pathways is useful to better distinguish the relative contributions of structural and electronic changes on magnetism, in addition to the relative effects of direct exchange and superexchange.

Because Cl substitution induces an expansion along the *b*-axis, the reduced magnitude of $J_3$ cannot be explained by direct exchange, and must instead be rationalized by the shift in the Cr–$S_1$–Cr bond angle towards 180°, which enhances the AFM contributions in the superexchange pathway (**table s4** and **figure s16**). Similarly, because Cl substitution has little effect on the $d_{Cr\text{–}Cr}$ relevant to $J_2$, direct exchange is unlikely to contribute strongly to changes in $J_2$. Surprisingly, while $J_2$ is calculated to become more strongly FM, the Cr–S–Cr bond angles relevant to $J_2$ increase away from 90° (**table s4**), suggesting that electronic, rather than structural modifications, must drive the changes in magnetic exchange. Here, we propose that the reduced covalency of the Cr–Cl interaction (compared to Cr–Br) leads to an increase in the Cr–S bond covalency (indicated by a reduction in $d_{Cr\text{–}S_2}$), which enhances the magnitude of the $J_2$ superexchange. Further, this change in the Cr–halogen covalency helps explain the changes in magnetic anisotropy with Cl substitution. Our experimental and computational data support large reductions in the magnetic anisotropy energy when Cl is substituted for Br (**table s4**), in line with previously predicted results[56]. The combined effects of reduced Cr–halogen covalency and smaller spin–orbit coupling for Cl compared to Br should dramatically weaken magnetocrystalline anisotropy in these materials, which is largely derived from anisotropic exchange interactions mediated by the halogens.

Intriguingly, a comparison of the Curie-Weiss analyses performed at the highest pressure (1.39 GPa – **figure s5**) and at the highest Cl substitution (Cl-67 – **figure s27**) reveal nearly identical changes in the $\theta_W$, implying similar changes in the overall magnitude of the intralayer FM exchange. However, the effects on the magnetic ordering temperature are much more dramatic for Cl substitution (-43.5 K vs. CrSBr) compared to pressure (-16.6 K at 1.39 GPa vs. ambient *P*), indicating that other factors play a key role in the magnetic ordering temperature of CrSBr and its analogues. The presence of interlayer frustration in the Cl-substituted compounds may partly explain the reduced critical temperatures, but the small magnitude of interlayer exchange compared to the intralayer exchange suggests this effect should play a small role in dictating the ordering temperature. Instead, we propose that the reduced magnetic anisotropy between the *a*- and *b*-axes in the alloys suppresses the magnetic ordering temperature. An intermediate magnetic regime with short-range FM correlations has been observed previously in CrSBr, and these results could support claims that this regime hosts 2D-XY-like behavior (**figure s32**)[42, 43], motivating further study of the magnetism of the mixed-halogen compounds at the 2D limit. More broadly, the effects of anisotropy observed here indicate that strong uniaxial anisotropy is required to maximize magnetic ordering temperatures for in-plane, orthorhombic 2D magnets and that 2D-XY-like magnetic regimes may be accessible outside of materials with high rotational symmetry.

In summary, we have demonstrated two routes to tune the magnetic properties of the layered semiconductor CrSBr: hydrostatic pressure and halogen substitution. Both strategies are found to reduce the magnetic ordering temperature. The combined experimental and computational analyses suggest a complex interplay between structural effects, weakly modulating both direct exchange and superexchange pathways, and changes to the Cr–halogen covalency, which strongly affects the strength of the FM superexchange. Additionally, in the Cl-substituted phases, we have identified a strong suppression of the anisotropy energy between the two in-plane axes, while maintaining a large anisotropy for the spins to remain in-plane. Preliminary optical and exfoliation experiments indicate that these Cl-substituted analogues retain the semiconducting properties and ambient stability of the parent CrSBr phase, motivating further characterization of the coupling between magnetism and optical, electronic, and structural properties across the series. More generally, these results highlight that the CrSBr family of 2D magnets offers the ability to chemically or mechanically control magnetic coupling and anisotropy, similar to the more thoroughly studied chromium trihalide family. The enhanced tunability of the interlayer coupling, improved stability in ambient conditions, and semiconducting transport properties strongly motivate the incorporation of CrSBr and its analogues into functional 2D spintronic devices. Further modulation of the properties of CrSBr through chalcogen alloying or iodine-substitution can expand upon the rich phase space of these materials, which includes diverse magnetic ground states (FM, AFM, spin glass) and spans a wide range of ordering temperatures, with the additional possibility to access 2D-XY magnetic phases hosting topological magnetic vortices. Finally, the demonstration that hydrostatic pressure, in addition to uniaxial strain, can tune the magnetism of these materials post-synthetically provides a new handle to modify the magnetic properties both in the bulk and at the 2D limit. These chemical and mechanical tools make the CrSBr family an ideal platform for future developments in 2D magnetism and spintronic devices.

**Supporting Information:**

Supporting information is available from the Wiley Online Library or from the authors.

**Acknowledgments:**

E.J.T. and D.G.C. contributed equally to this work.

We thank Dr. Yue Meng and Rich Ferry for assistance with DAC assembly and high-pressure X-ray diffraction measurements. Research on tunable vdW magnetic semiconductors was supported as part of Programmable Quantum Materials, an Energy Frontier Research Center funded by the U.S. Department of Energy (DOE), Office of Science, Basic Energy Sciences (BES), under award DE-SC0019443. Synthesis and structural characterization of mixed halide compounds was supported by the Columbia MRSEC on Precision-Assembled Quantum Materials (PAQM) under award number DMR-2011738. The first-principles calculations are mainly supported by NSF MRSEC DMR-1719797. T.C. acknowledges support from the Micron Foundation. Computational resources were provided by HYAK at the University of Washington. High pressure powder X-ray diffraction measurements were performed at beamline 16-ID-B at HPCAT (Sector 16), Advanced Photon Source (APS), Argonne National Laboratory. HPCAT operations are supported by DOE-NNSA's Office of Experimental Sciences. Use of the COMPRES-GSECARS gas loading system at the APS was supported by COMPRES under NSF Cooperative Agreement EAR -1606856 and by GSECARS through NSF grant EAR-1634415 and DOE grant DE-FG02-94ER14466. The Advanced Photon Source is a U.S. Department of Energy (DOE) Office of Science User Facility operated for the DOE Office of Science by Argonne National Laboratory under Contract No. DE-AC02-06CH11357. C.-Y.H. is supported by the Taiwan-Columbia Fellowship funded by the Ministry of Education of Taiwan and Columbia University. The PPMS used to perform magnetic susceptibility measurements was purchased with financial support from the NSF through a supplement to award DMR-1751949. The Columbia University Shared Materials Characterization Laboratory (SMCL) was used extensively for this research. We are grateful to Columbia University for the support of this facility.

**Conflict of Interest:**

The authors declare no conflict of interest.

Supporting Information

**Designing magnetic properties in CrSBr through hydrostatic pressure and ligand substitution**

*Evan J. Telford[Δ], Daniel G. Chica[Δ], Kaichen Xie, Nicholas S. Manganaro, Chun-Ying Huang, Jordan Cox, Avalon H. Dismukes, Xiaoyang Zhu, James P. S. Walsh, Ting Cao, Cory R. Dean, Xavier Roy*, Michael E. Ziebel**

Evan J. Telford, Daniel G. Chica, Chun-Ying Huang, Jordan Cox, Avalon H. Dismukes, Prof. Xiaoyang Zhu, Prof. Xavier Roy, Michael E. Ziebel
Department of Chemistry, Columbia University, New York, NY, USA
E-mail : xr2114@columbia.edu, mez2127@columbia.edu

Evan J. Telford, Prof. Cory R. Dean
Department of Physics, Columbia University, New York, NY, USA

Kaichen Xie, Prof. Ting Cao
Department of Materials Science and Engineering, University of Washington, Seattle, WA, USA

Nicholas S. Manganaro, Prof. James P. S. Walsh
Department of Chemistry, University of Massachusetts Amherst, Amherst, MA, USA

Δ These authors contributed equally to this work

**Synthesis of CrSBr:**
Large single crystals of CrSBr were grown using a chemical vapor transport reaction described in ref [1].

**Synthesis of $CrSBr_{1-x}Cl_x$:**
The synthesis of Cl-alloyed CrSBr was achieved using a modified reaction of the pure CrSBr reaction. Chromium metal (99.94%, -200 mesh, Alfa Aesar), sulfur pieces (99.9995%, Alfa Aesar), chromium(III) chloride (anhydrous, 99.9%, Thermo scientific), and bromine (99.99%, Aldrich) were used as received. Chromium(III) bromide was synthesized as described in ref [1]. In a typical reaction, a slightly off stoichiometric ratio of the reagents with a total mass of 1 gram (see table below for ratio of reagents for each particular Cl-alloyed CrSBr sample) were loaded into a 12.7 mm o.d., 10.5 mm i.d. fused silica tube which was sealed to a length of 20 cm. The tube was subjected to the following heating profile using a computer controlled two zone horizontal tube furnace: *Source side*: Heat to 800°C in 24 hours, soak for 48 hours, heat to 875°C in 12 hours, soak for 72 hours and then water quench. *Sink side*: Heat to 875°C in 24 hours, soak for 48 hours, heat to 800°C in 12 hours, soak for 72 hours and then water quench. ***Caution!*** When quenching the reaction ensure proper PPE is used including a full-face shield, fire resistant lab coat, and a blast shield.

| Target composition | $Cr:S:CrCl_3:CrBr_3$ target ratio | Single crystal composition |
|---|---|---|
| CrSBr | 2/3:1:0:1/3 | CrSBr |
| $CrSCl_{1/8}Br_{7/8}$ | 2/3:1:1/24:7/24 | $CrSCl_{0.11}Br_{0.89}$ |
| $CrSCl_{1/3}Br_{2/3}$ | 2/3:1:1/9:2/9 | $CrSCl_{0.27}Br_{0.73}$ |
| $CrSCl_{1/2}Br_{1/2}$ | 2/3:1:1/6:1/6 | $CrSCl_{0.41}Br_{0.59}$ |
| $CrSCl_{5/8}Br_{3/8}$ | 2/3:1:5/24:1/8 | $CrSCl_{0.57}Br_{0.43}$ |
| $CrSCl_{3/4}Br_{1/4}$ | 2/3:1:1/4:1/12 | $CrSCl_{0.67}Br_{0.33}$ |

Note: All compositions were synthesized through chemical vapor transport (CVT) reactions using a stoichiometric amount of chromium(III) bromide/chloride, sulfur, and chromium. CVT reactions rely on all the elements having large enough partial pressures for effective mass transport through the formation of volatile transport effective species which were generated *in situ* at the crystal growth temperatures (850C - 950C). The temperatures used in the synthesis allowed for both halogens species to transport effectively and be incorporated into the final product; though, the final composition of the product was typically deficient in chloride (i.e., the nominal ratio of chlorine to bromine used in the synthesis was greater than the ratio derived from SCXRD and EDX). CVT reactions with higher chlorine concentrations were attempted though only resulted in the deposition of $Cr(Cl/Br)_3$ and $Cr_2S_3$ phases on the sink side limiting the highest chlorine alloy level to Cl-67. Note that the original synthesis[2] for Cl-alloyed CrSBr required the use of $S_2Cl_2$ and $S_2Br_2$. Because these reagents are liquid, using the original method limits the precise control of the stoichiometry compared to solids which can be mass accurately. Additionally, the original synthesis incorporated only 1/3 Cl onto the Br sites while the method described in this work can incorporate double the amount of Cl.

**Powderization of CrSBr crystals for magnetometry measurements under pressure:**
CrSBr was powderized through the following process: large crystals of CrSBr were placed in a thin porcelain crucible along with enough liquid $N_2$ to fully submerge the crystals. The crystals were ground with a thermally equilibrated pestle for 5 mins. The material was rinsed with acetone to remove residual moisture from condensation.

**Determination of applied hydrostatic pressure for magnetometry measurements under pressure:**
Since the superconducting critical temperature ($T_C$) of Pb is well-known to linearly depend upon the applied hydrostatic pressure at a rate of $dT_C/dP$ = 0.379 K·GPa$^{-1}$[3], we can use the measured $T_C$ of Pb to determine the applied hydrostatic pressure on CrSBr. The Pb plus CrSBr sample is first zero-field cooled below the transition to 6 K, then the magnetic susceptibility ($\chi$) versus temperature ($T$) is measured with a small measuring field of 5 Oe (such that the measuring field is much less than the zero-temperature upper critical field[4], which for lead is 800 Oe). $\chi$ versus $T$ is measured at a rate of 0.05 K/min to ensure the transition is precisely resolved and traces with increasing and decreasing $T$ were measured to check for measurement precision. The Pb $T_C$ is extracted by finding the condition where $\chi = 0.5\chi_N$ (where $\chi_N$ is the susceptibility in the normal state) and correlated to the measured pressure-cell compression.

**Vibrating sample magnetometry under hydrostatic pressure:**
All vibrating sample magnetometry was conducted on a Quantum Design PPMS DynaCool system using the commercially-available HMD high pressure cell. Multiple single CrSBr crystals were selected, and powderized in liquid nitrogen using a mortar and pestle. Before and after the VSM measurements, PXRD was used to confirm there was no significant change in structure upon powderizing or after applying maximum pressure. The powder was then combined with Daphne 7373 oil and a ~1-2 mm long wire of Pb in a Teflon capsule was inserted into the pressure cell. The variable temperature scans and field-dependent magnetic susceptibility curves for each pressure were measured during the same measurement cycle. The measurements performed at different pressures were done sequentially with increasing pressure (from zero applied pressure up to the maximum achievable pressure). After the final maximum pressure measurement, the capsule containing the CrSBr powder, Daphne 7373 oil, and the Pb manometer was removed, fixed to a brass paddle with GE varnish, and re-measured as a consistency check of the zero-pressure measurement.

**Vibrating sample magnetometry on $CrSBr_{1-x}Cl_x$:**
All vibrating sample magnetometry was conducted on a Quantum Design PPMS DynaCool system. For each stoichiometry, a pristine single $CrSBr_{1-x}Cl_x$ crystal was selected and attached to a quartz paddle using GE varnish (which was cured at room temperature under ambient conditions for 30 minutes) and oriented with the *a*-, *b*-, or *c*-axis parallel to applied field direction. The same crystal was used for all axial-orientated measurements. The variable temperature scans and field-dependent magnetic susceptibility curves for each axis were measured during the same measurement cycle. Between axial-oriented measurements, the crystal was removed using a 1:1 ethanol/toluene solution, dried in air and then reoriented and reattached using GE varnish.

**Ac magnetometry on $CrSBr_{1-x}Cl_x$:**
All ac magnetometry was conducted on a Quantum Design PPMS DynaCool system with the ACMSII module. For each measured stoichiometry, a pristine single $CrSBr_{1-x}Cl_x$ crystal was selected and attached to a quartz paddle using GE varnish (which was cured at room temperature under ambient conditions for 30 minutes) and oriented with the *a*- or *b*-axis parallel to the applied magnetic field. An ac magnetic field excitation of 4 Oe was used for all measurements. The variable temperature and frequency-dependent magnetic susceptibility curves for each axis were measured during the same measurement cycle.

**Ambient-pressure powder X-ray diffraction:**
Powder diffraction patterns were collected on a Malvern Panalytical Aeris diffractometer with a Cu Kα x-ray source energized to 40 kV and 15 mA. The X-ray beam was filtered with a Niβ filter. The LN-powderized sample of CrSBr was mounted on a Si- zero background holder which was spun during the collection to reduce preferred orientation.

**Single-crystal X-ray diffraction:**
Single crystal diffraction measurements were collected on $CrSBr_{1-x}Cl_x$ crystals using an Agilent Supernova single crystal diffractometer. The crystals were mounted onto a MiTeGen MicroLoops™ holder with paratone oil. The X-ray source was a Mo Kα micro-focus energized to 50 kV and 0.8 mA. The collection temperature was maintained at 250 K using an Oxford instruments nitrogen cryostat. The data collection, integration, and reduction were performed using the Crysalis–Pro software suite. The crystal structure was solved and refined using ShelXT and ShelXL respectively.

**Details of diamond anvil cell (DAC) assembly.**
We used Boehler–Almax diamond anvils with 300 μm culets set in tungsten carbide seats with a conical aperture of 80°. The anvils and seats were loaded into DacTools iBX-80 type cells. A stainless-steel gasket with a starting thickness of 250 μm was pre-indented to a thickness of ~40 μm. A sample space with a diameter of ~200 μm was then created in the center of the indented gasket via electro-discharge machining using a Boehler μDrill with a copper wire electrode.

**High pressure powder X-ray diffraction measurements.**
To reduce texture effects in powder X-ray diffraction measurements, single crystals of CrSBr were first cooled to 77 K in liquid nitrogen and then ground with a mortar and pestle. The resulting powder was sieved to remove large, unground crystals. The sieved powder was further ground between two glass slides prior to loading in the diamond anvil cell.

The sample chamber prepared as described above was loaded with CrSBr powder, a small piece of gold foil to serve as a pressure calibrant during diffraction measurements, and two ruby microspheres (BETSA®) to serve as a pressure calibrant during gas loading. A representative photograph of one of the loaded cells is shown in **figure s11**. The cell was subsequently loaded with neon as the pressure transmitting medium using the COMPRES gas loading system as GSECARS, at the Advanced Photon Source at Argonne National Laboratory[5].

High pressure powder X-ray diffraction experiments were conducted at beamline 16-ID-B, within HPCAT at the Advanced Photon Source (APS). High intensity monochromatic synchrotron radiation with a fixed wavelength of 0.406626 Å was used as the source in all diffraction measurements. The cell was loaded into a diaphragm gas membrane assembly, which enables diffraction measurements over very small pressure increments (~0.1 GPa). At each pressure step, separate diffraction images were collected without rotation on the CrSBr sample and the Au foil to enable determination of lattice parameters and sample-space pressure, respectively. Diffraction images were masked and integrated using the Dipotas 0.5.1 software package to produce the corresponding 1D diffraction patterns[6].

**Analysis of Powder X-ray Diffraction Data**.
For each pressure step, the cell pressure was obtained by comparison of the lattice parameters of the Au foil with the established equation of state[7]. Powder X-ray diffraction data were then analyzed using the GSAS-II software package[8]. Due to the weak intensity of the (00*l*) reflections and the possible overlap of the (011) and (002) reflections, we observed that lattice parameters obtained using the Pawley method were highly sensitive to the initial parameters used in the refinement. To obtain reasonable initial parameters, we extracted the estimated *b*-lattice parameter by inspection of the (020) reflection, and subsequently estimated *a*- and *c*-lattice parameters by inspection of the (110) and (011) reflections, respectively. Using these lattice parameters as the initial values, we then fit the patterns over the $2\theta$ range 3° to 23° using the Pawley method to extract accurate unit cell parameters at each pressure. We note that it was necessary to constrain the b-axis lattice parameter during initial refinements of the background, line shape, and *a*- and *c*- axis lattice parameters to obtain reasonable fits of the (020) reflection.

We then used the software package EoSFit7[8] to fit the unit cell volume as a function of pressure. We used a third-order Birch–Murnaghan equation of state to fit the data[9, 10]:

$$P(V) = \frac{3B_0}{2}\left(\left(\frac{V_0}{V}\right)^{\frac{7}{3}} - \left(\frac{V_0}{V}\right)^{\frac{5}{3}}\right)\left\{1 + \frac{3}{4}(B_0' - 4)\left(\left(\frac{V_0}{V}\right)^{\frac{2}{3}} - 1\right)\right\}$$

Where $P$ is the pressure, $V$ is the unit cell volume, $V_0$ is the initial unit cell volume at ambient pressure, $B_0$ is the bulk modulus, and $B_0$' is the derivative of the bulk modulus with respect to pressure. A single equation of state was sufficient to fit the data at room temperature up to 3.5 GPa, suggesting no phase transition occurs in the pressure range where magnetic analyses were performed. A small anomaly is possibly observed in the *b*-axis lattice parameters near 0.6 GPa, though we attribute this anomaly to the necessary constraints applied to the *b*-axis lattice parameter during refinements, as described above.

**Scanning electron microscopy:**
Scanning electron micrographs were collected on a Zeiss Sigma VP scanning electron microscope (SEM) using a beam energy of 5 kV. Energy dispersive X-ray spectroscopy (EDX) of the CrSBr crystals was performed with a Bruker XFlash 6 | 30 attachment. Spectra were collected with a beam energy of 15 kV. Elemental compositions and atomic percentages were estimated by integrating under the characteristic spectrum peaks for each element using Bruker ESPRIT 2 software.

**Raman spectroscopy:**
Raman spectroscopy for all $CrSBr_{1-x}Cl_x$ single crystals was performed under ambient conditions in a Renishaw InVia™ micro-Raman microscope using a 532 nm wavelength laser. A 50x objective was used with a laser spot size of 2-3 μm. A laser power of ~2 mW was used with a grating of 2400 g/mm for all spectra. An acquisition time of 20s was used for each measurement. For each crystal, 5 independent spectra were acquired and averaged after subtracting a dark background. The dark background was a spectrum acquired with no laser excitation and the same acquisition parameters.

**Photoluminescence (PL) spectroscopy:**
PL measurements were carried out with a 450-nm continuous-wave (CW) laser with a power of 900 μW. The PL spectra were collected by a Princeton Instruments PyLoN-IR detector cooled with liquid nitrogen. All samples were prepared by exfoliating single crystals of $CrSBr_{1-x}Cl_x$ onto $SiO_2$/Si+ substrates passivated with 1-dodecanol. The exfoliation was done under inert conditions in an $N_2$ glovebox with < 1 ppm $O_2$ and < 1 ppm $H_2O$ content. Thin-bulk flakes were identified by optical microscopy and loaded into an Oxford Instruments Microstat HiRes2 cryostat inside the glovebox to avoid exposing the samples to air before measurements.

**Exfoliation:**
$CrSBr_{1-x}Cl_x$ flakes were exfoliated onto 285 nm $SiO_2$/Si+ substrates using mechanical exfoliation with Scotch® Magic™ tape[11, 12]. Before exfoliation, the substrates were cleaned with a gentle oxygen plasma to remove adsorbates from the surface and increase flake adhesion[13]. The exfoliation was done under inert conditions in an $N_2$ glovebox with < 1 ppm $O_2$ and < 1 ppm $H_2O$ content. Flake thickness was identified using optical contrast and then confirmed with atomic force microscopy.

**Atomic Force Microscopy:**
Atomic force microscopy was performed in a Bruker Dimension Icon® using OTESPA-R3 tips in tapping mode. Flake thicknesses were extracted using Gwyddion to measure histograms of the height difference between the substrate and the desired flake.

**Theoretical Calculations:**
Ab initio calculations of bulk CrSBr and CrSCl were performed using DFT implemented in the QUANTUM ESPRESSO package[14]. Norm-conserving pseudopotentials with a plane-wave energy cutoff of 85 Ry were employed. For structural optimization, the spin-polarized Perdew-Burke-Ernzerhof exchange-correlation functional was employed, with dispersion corrections within the D2 formalism[15] (PBE-D2) included to account for the van der Waals interactions. The structures were fully relaxed until the force on each atom was $< 0.005\ \mathrm{eV}\ \text{Å}^{-1}$. The calculated lattice constants for bulk CrSBr and CrSCl are 3.5 and 3.4 Å along the $a$ axis, respectively, and both 4.7 Å along the $b$ axis. The calculated interlayer distance for bulk CrSBr and CrSCl are 8 and 7.5 Å, respectively. For each hydrostatic pressure applied, the intra- and interlayer Heisenberg magnetic exchange couplings $J$ were calculated in $3 \times 3 \times 1$ and $3 \times 3 \times 2$ supercells respectively, by a four-state mapping method[16] within the local spin density approximation (LSDA). The Curie temperature was calculated using metropolis Monte Carlo (MC) methods implemented in the VAMPIRE package[17]. The critical exponent was determined by fitting the temperature dependent normalized magnetization $m(T)$ to the Curie-Bloch equation in the classical limit $m(T) = \left(1 - \frac{T}{T_C}\right)^{\beta}$. The saturation fields along different axes were extracted based on the Heisenberg model $H = H_0 + H_{\mathrm{inter}} - g\mu_B \sum_i h \cdot \mathbf{S}_i$, where $H_{\mathrm{inter}} = \sum_{i \in t, j \in b} J_{\mathrm{inter},ij} \mathbf{S}_i \cdot \mathbf{S}_j$ with $t$ and $b$ denote the top and bottom layers in a unit cell, $h$ represents the external magnetic field. The ground state energy differences between the FM and AFM states ($E^{\mathrm{FM}} - E^{\mathrm{AFM}}$) under different hydrostatic pressure were calculated with spin-orbit coupling (SOC) taken into account within LSDA, based on the structures revealed by PBE-D2.

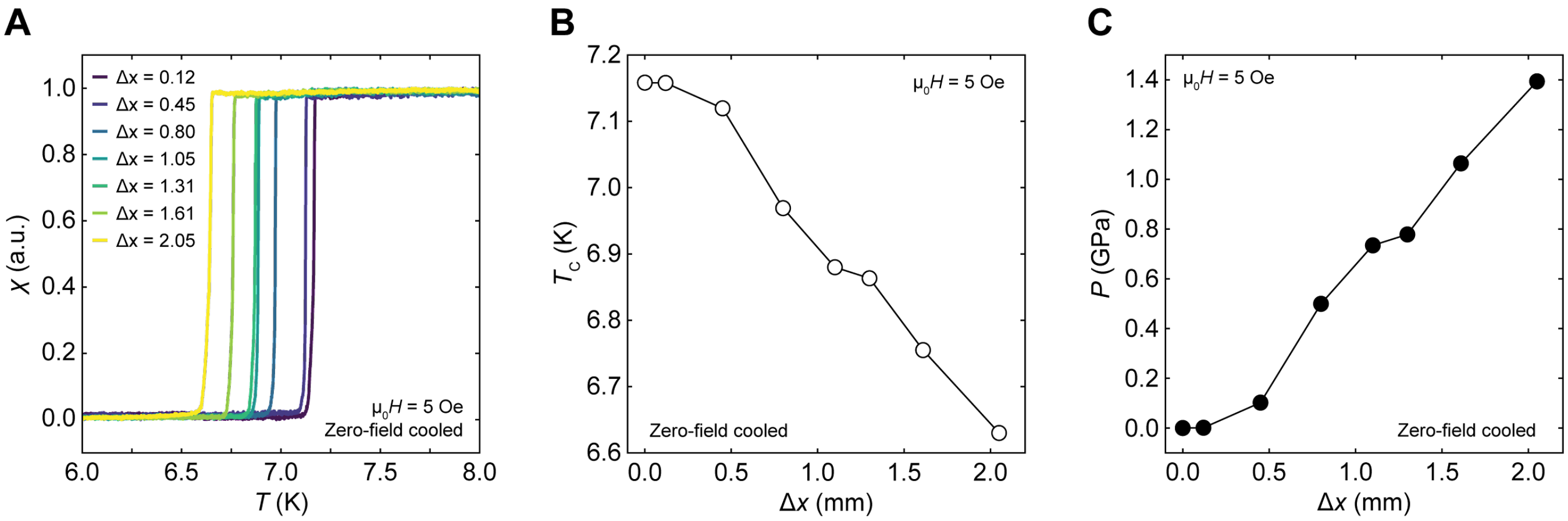

**Figure s1: Determination of applied pressure versus cell compression. A)** Magnetic susceptibility versus temperature across the superconducting transition for the lead manometer for various applied pressures. The corresponding cell compression is given in the inset. All traces were offset and normalized for clarity. A measuring field of 5 Oe was used. **B)** Extracted superconducting transition temperature ($T_C$) of lead versus cell compression. $T_C$ was defined as the temperature at which the magnetic susceptibility was 50% of the normal-state susceptibility. **C)** Corresponding applied pressure versus cell compression. The applied pressure was determined from the well-known $dT_C/dP$ = 0.379 K/GPa for lead[3].

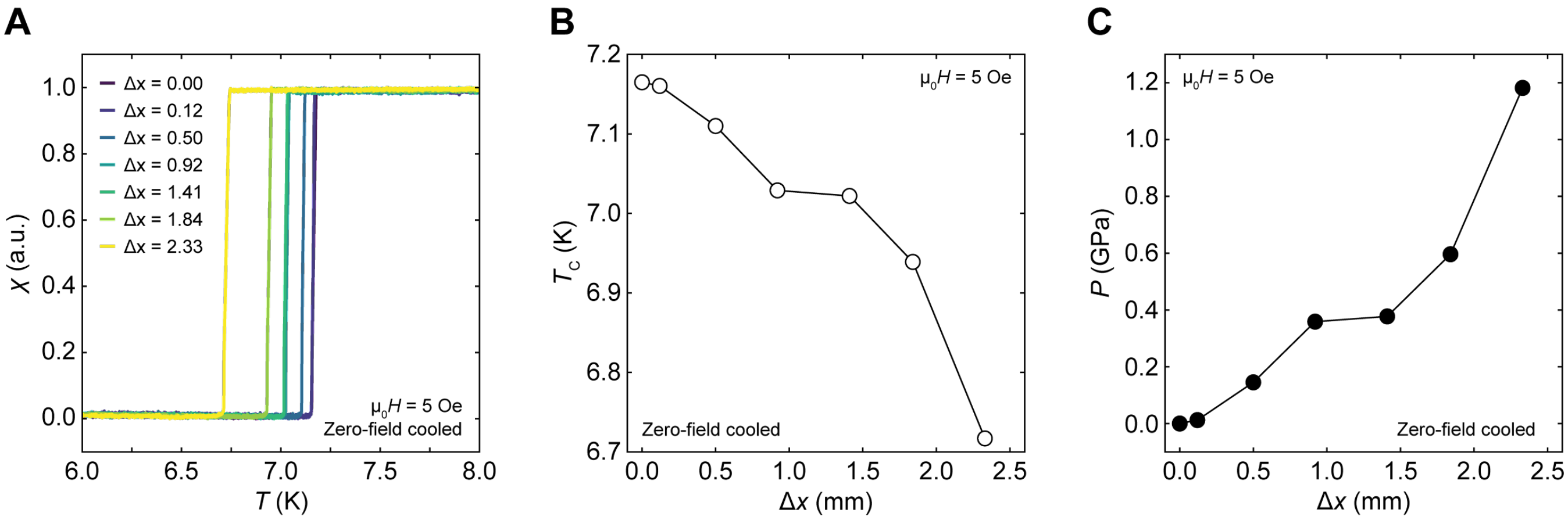

**Figure s2: Determination of applied pressure versus cell compression for measurement run 2. A)** Magnetic susceptibility versus temperature across the superconducting transition for the lead manometer for various applied pressures. The corresponding cell compression is given in the inset. All traces were offset and normalized for clarity. A measuring field of 5 Oe was used. **B)** Extracted superconducting transition temperature ($T_C$) of lead versus cell compression. $T_C$ was defined as the temperature at which the magnetic susceptibility was 50% of the normal-state susceptibility. **C)** Corresponding applied pressure versus cell compression. The applied pressure was determined from the well-known $dT_C/dP$ = 0.379 K/GPa for lead[3].

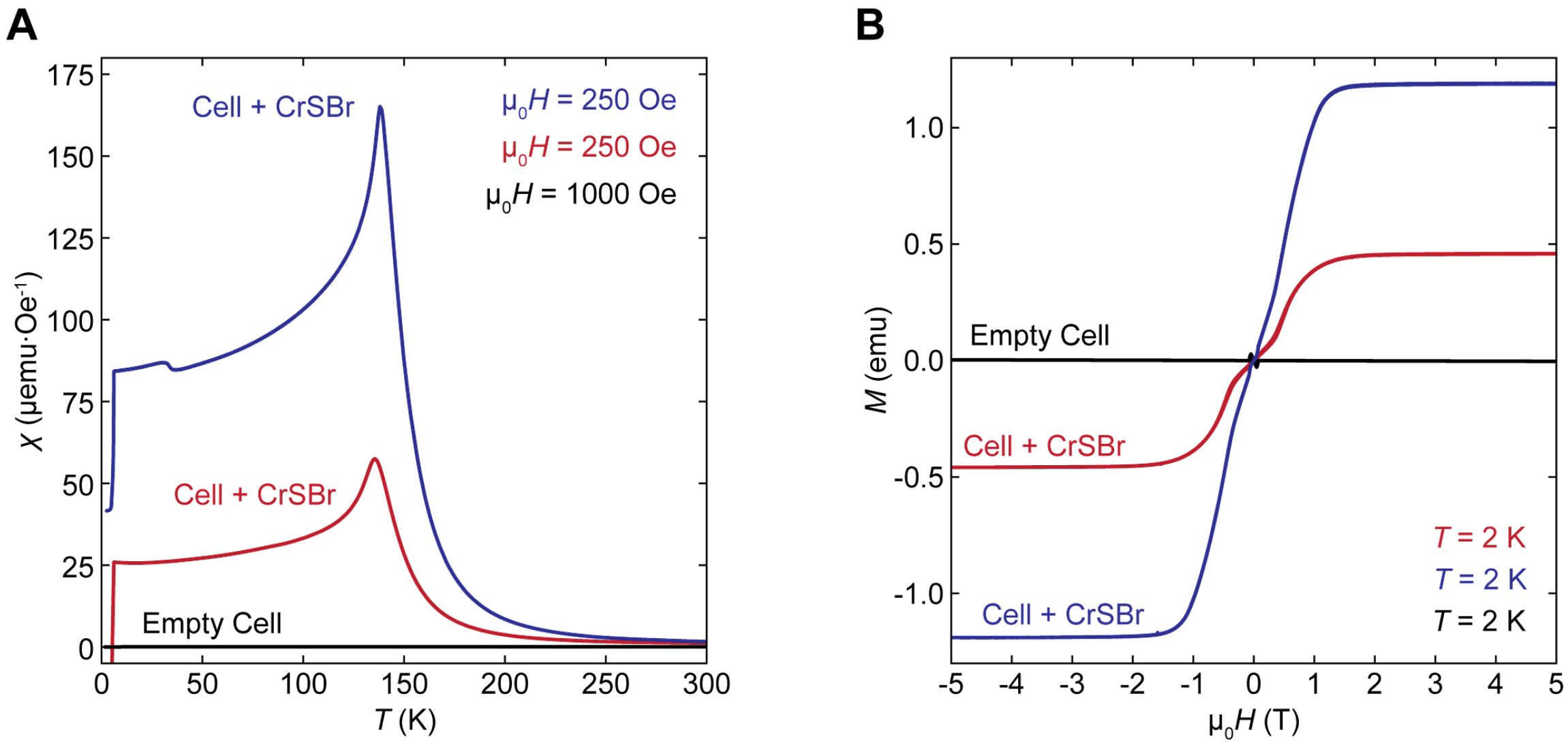

**Figure s3: Measurements of the Be-Cu pressure cell background. A)** Zero-field-cooled magnetic susceptibility of the empty pressure cell (solid black trace) and the pressure cell loaded with CrSBr powder (solid blue trace and solid red trace for both batches). The corresponding measuring fields are given in the inset. **B)** Magnetization versus magnetic field at 2 K for the empty pressure cell (solid black trace) and the pressure cell loaded with CrSBr powder (solid blue trace and solid red trace for both batches).

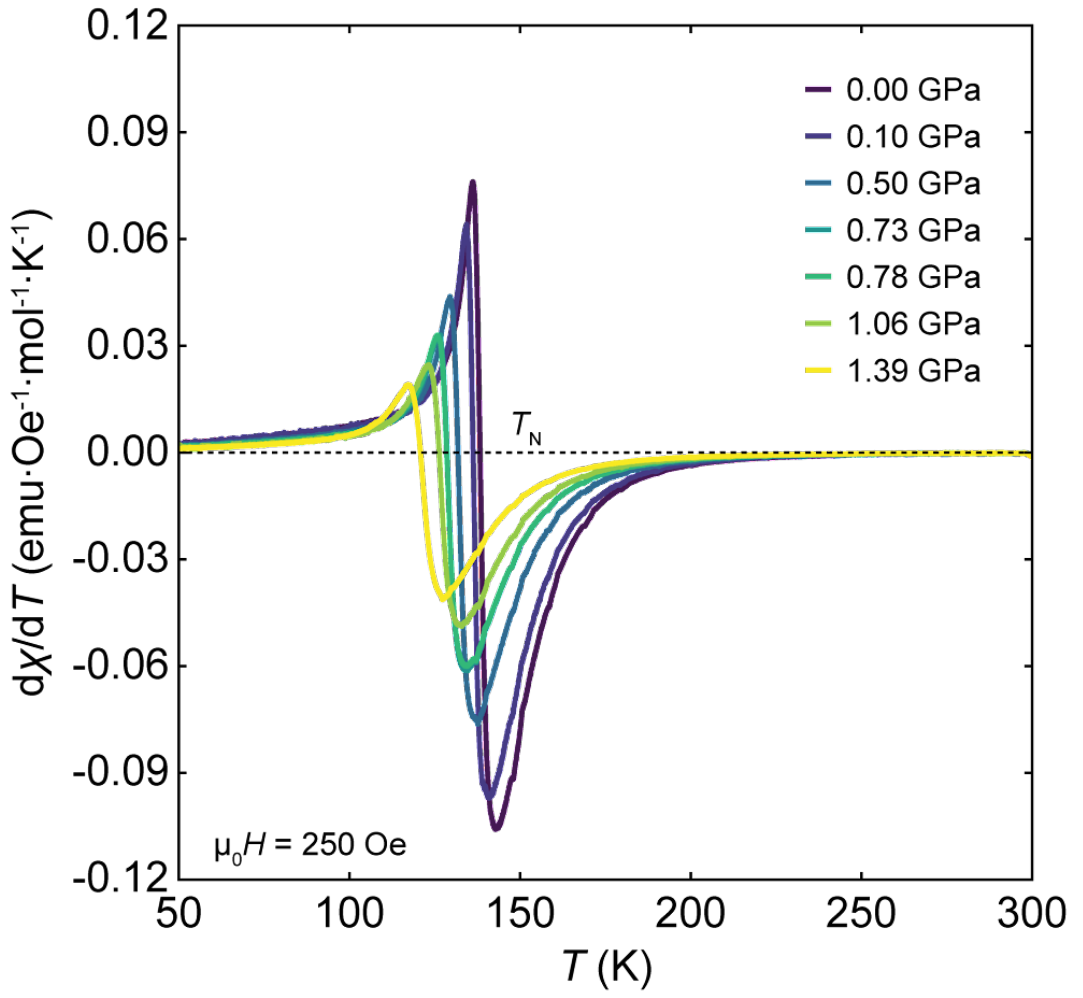

**Figure s4: Determination of $T_N$ for measurement run 1.** Derivative of the magnetic susceptibility versus temperature for various applied pressures. The corresponding cell compression is given in the inset. The zero-crossing denotes $T_N$.

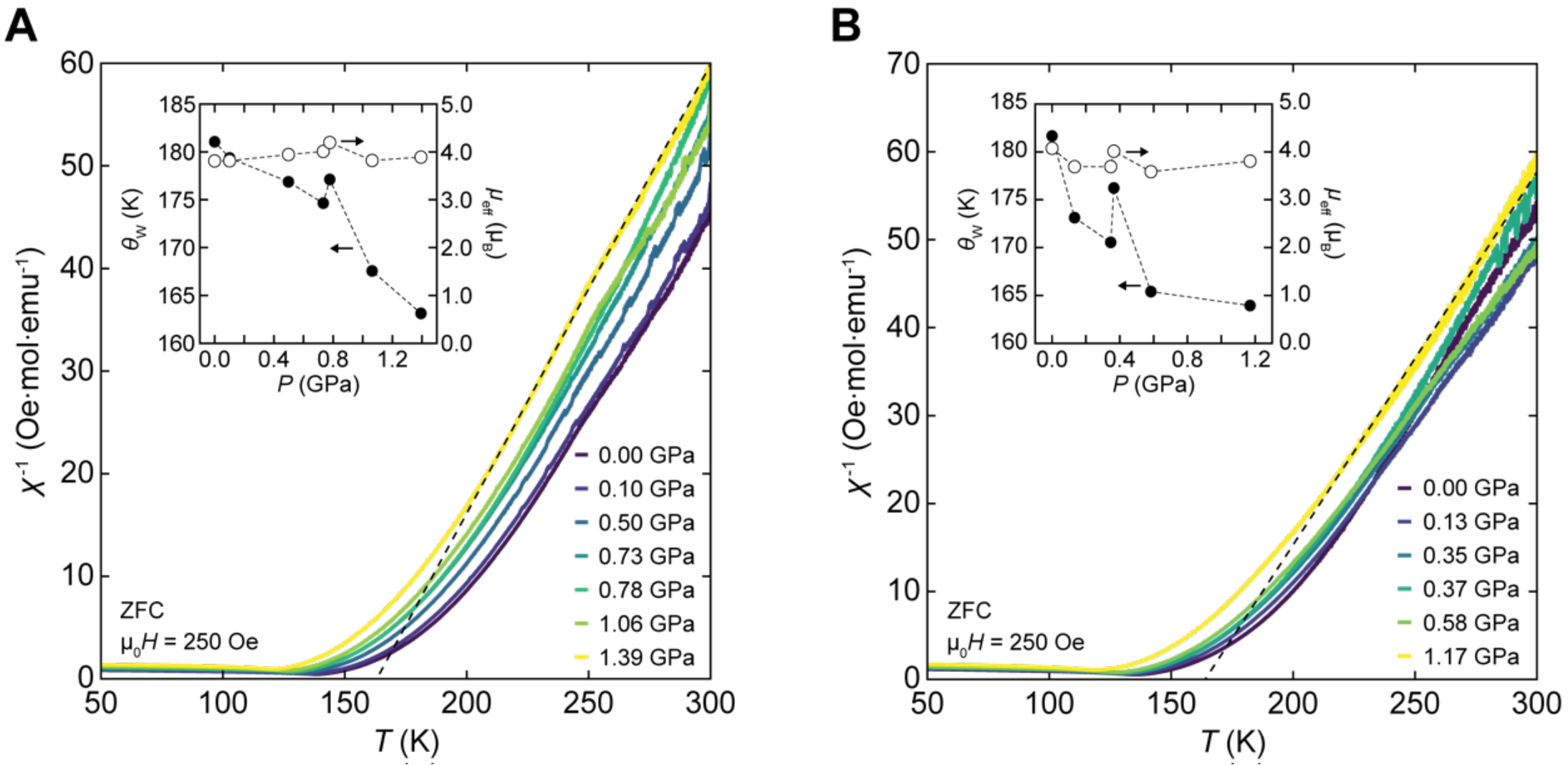

**Figure s5: Curie-Weiss analysis versus hydrostatic pressure. A, B)** Inverse susceptibility versus temperature of CrSBr sample 1 (**A**) and sample 2 (**B**) at various applied pressures. The pressure values for each trace are denoted in the lower-right inset. The dashed black lines are linear fits to the highest-applied-pressure trace. Top-left insets plot the extracted Weiss temperatures (solid black dots) and effective $Cr^{3+}$ magnetic moments (hollow black dots) versus hydrostatic pressure.

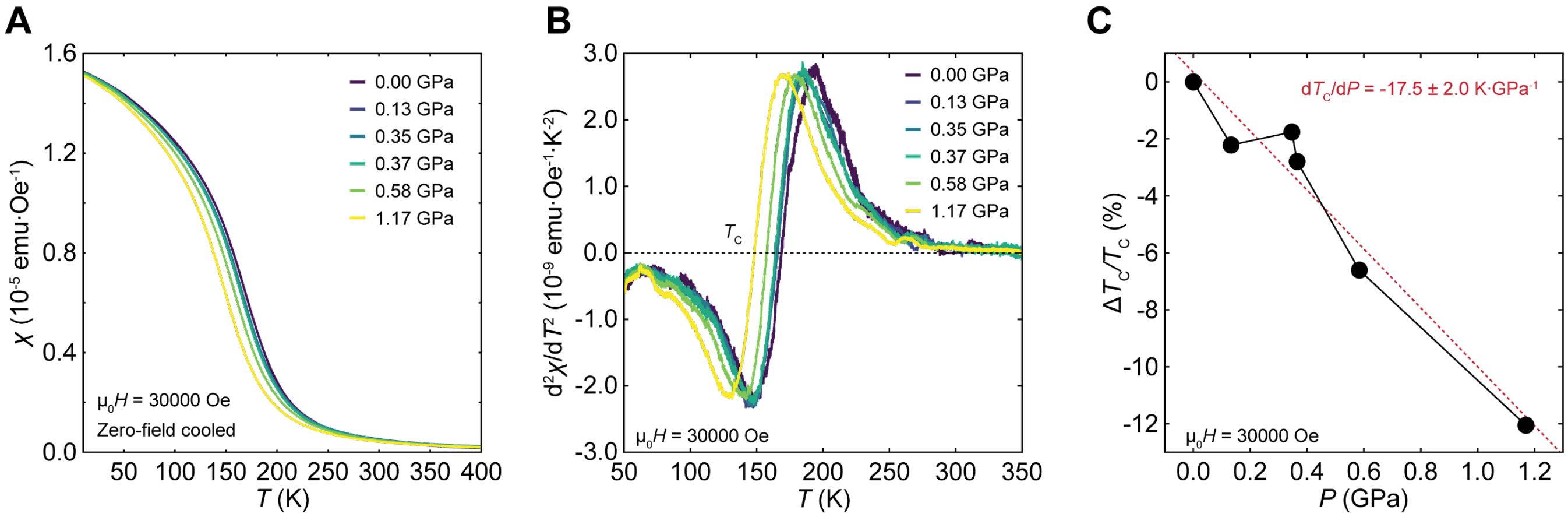

**Figure s6: High-field magnetic properties of CrSBr from measurement run 2. A)** Zero-field-cooled magnetic susceptibility versus temperature for various applied pressures. The corresponding pressure is given in the inset. A measuring field of 30000 Oe was used. **B)** Second derivative of the magnetic susceptibility versus temperature for various applied pressures. The corresponding pressure is given in the inset. The zero-crossing denotes $T_C$. **C)** Extracted percentage change in $T_C$ versus applied pressure. The slope of $T_C$ versus pressure is given in the inset and the corresponding linear fit to the data is denoted by a red dashed line.

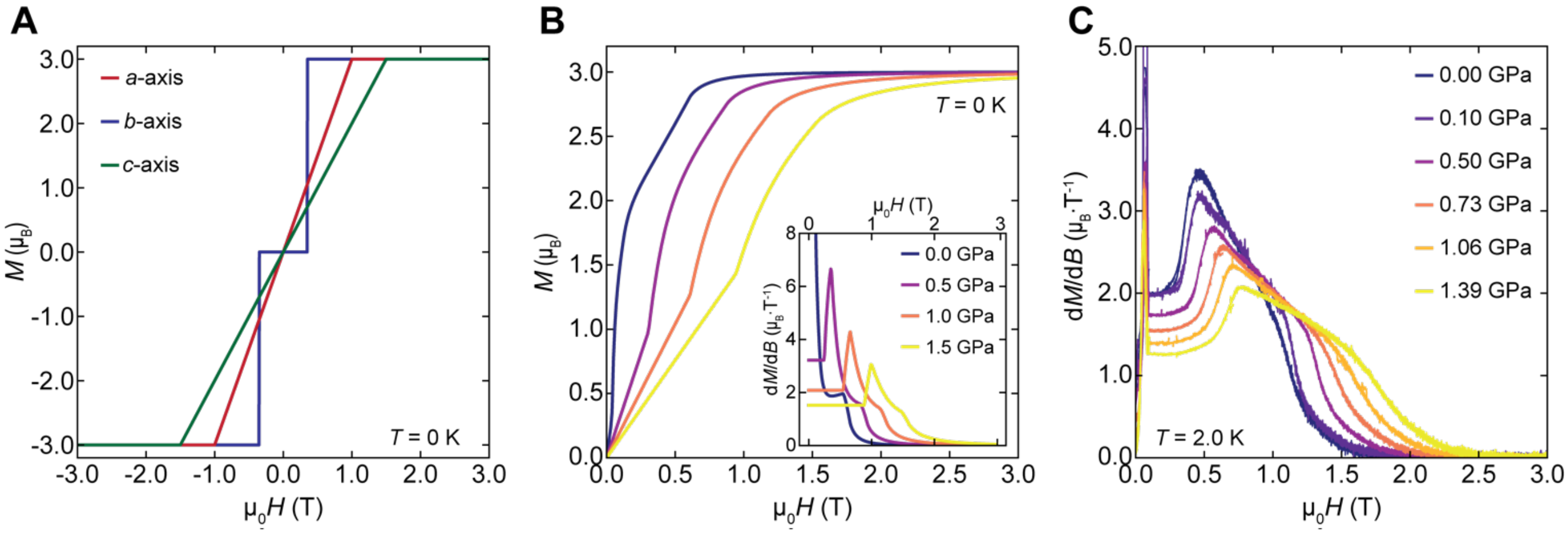

**Figure s7: Calculated magnetization versus field curves for CrSBr powder. A)** Expected magnetization versus magnetic field for bulk CrSBr at zero temperature for magnetic fields oriented along the *a*-axis (solid red line), *b*-axis (solid blue line), and the *c*-axis (solid green line). **B)** Calculated magnetization versus magnetic field under various hydrostatic pressures for a CrSBr powder in which the orientation of the crystals is random. The *a*-, *b*-, and *c*-axis saturation fields at each pressure are taken from **table s1**. Inset: Calculated derivative of the magnetization versus magnetic field under various pressures. Features are seen at each axial saturation field (a sharp kink at the *b*-axis saturation field, a peak at the *a*-axis saturation field, and a shoulder at the *c*-axis saturation field). **C)** Experimental derivative of the magnetization versus magnetic field at various applied pressures (taken from the data in **figure 2B**). There is close agreement between the calculated and experimental *M* vs $\mu_0 H$ curves.

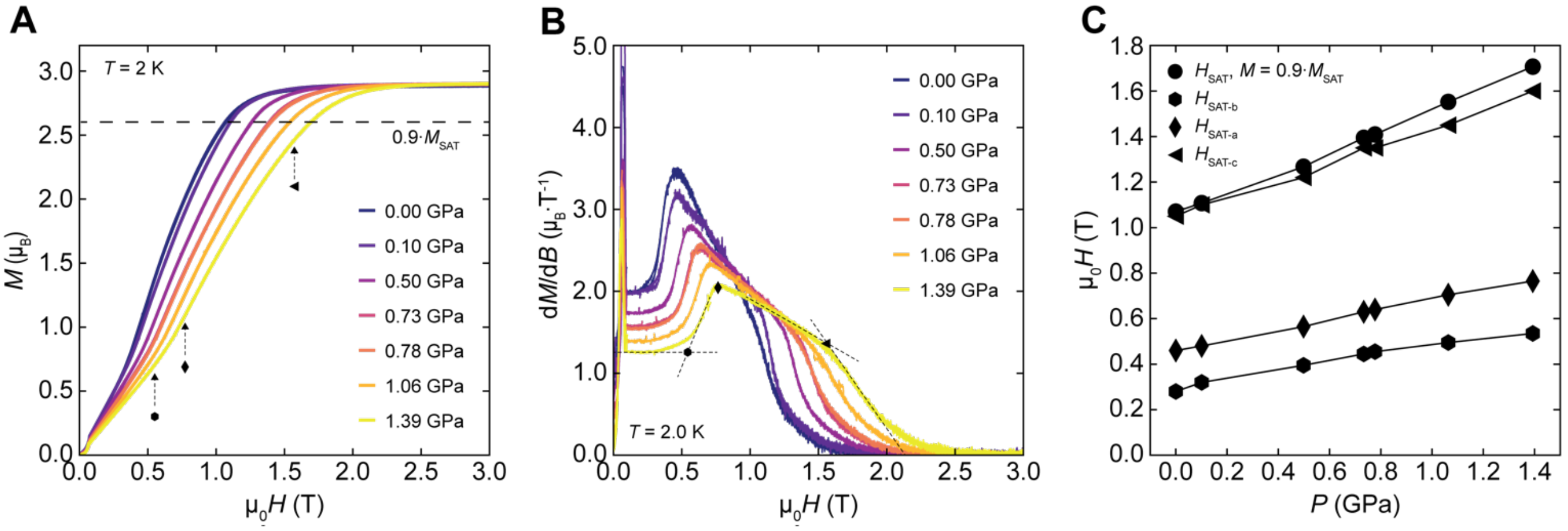

**Figure s8: Extracted saturation fields versus hydrostatic pressure. A, B)** Magnetization (*M*) (**A**) and derivative of the magnetization (d*M*/d*B*) (**B**) versus applied magnetic field ($\mu_0 H$) at 2 K for various *P*. Due to the pressure-cell preparation, $\mu_0 H$ is randomly oriented along all crystal axes. In both (**A**) and (**B**), features associated with the *a*-, *b*-, and *c*-axis saturation fields for the 1.39 GPa trace are denoted by black diamonds, black hexagons, and black triangles, respectively. The overall saturation field ($H_{SAT}$) is defined as the $\mu_0 H$ at which *M* is 90% of the saturation *M* (denoted by a black dashed line in (**A**)). **C)** Extracted saturation fields versus *P*.

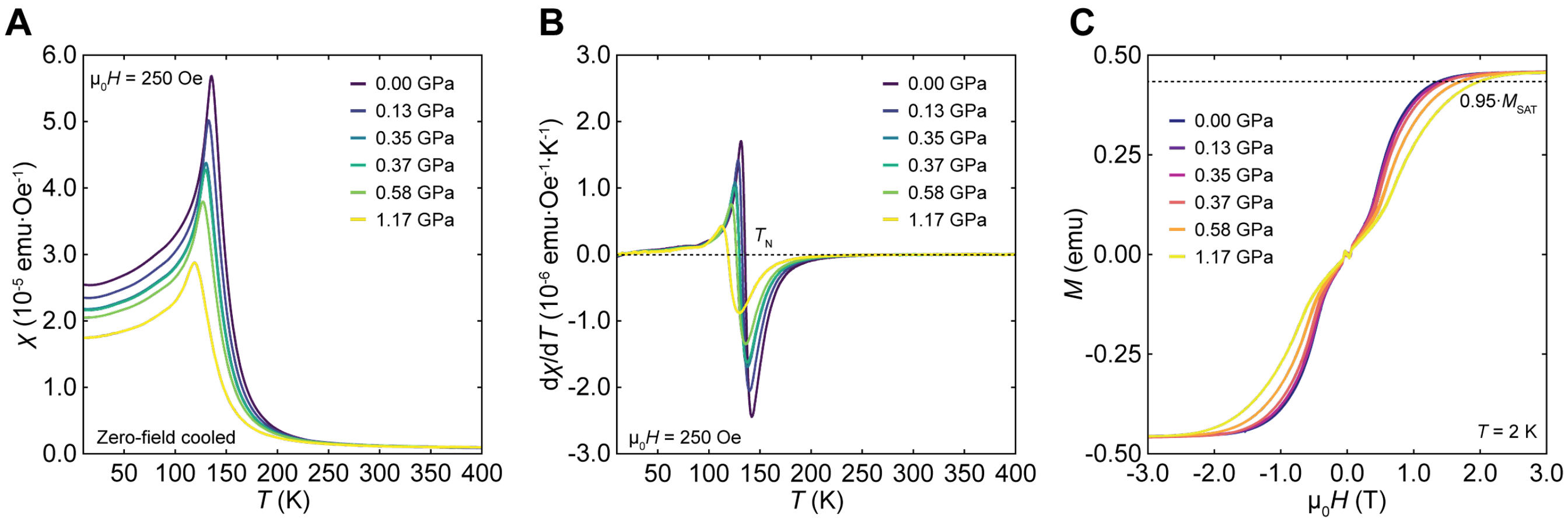

**Figure s9: Low-field magnetic properties of CrSBr from measurement run 2. A)** Zero-field-cooled magnetic susceptibility versus temperature for various applied pressures. The corresponding pressure is given in the inset. A measuring field of 250 Oe was used. **B)** Derivative of the magnetic susceptibility versus temperature for various applied pressures. The corresponding pressure is given in the inset. The zero-crossing denotes $T_N$. **C)** Magnetization versus magnetic field at 2 K for various applied pressures. Due to the pressure cell preparation, the magnetic field is randomly oriented along all crystal axes. The corresponding pressure is given in the inset. The saturation field is defined as the magnetic field at which the magnetization is 95% of the saturation magnetization (denoted by a black dashed line).

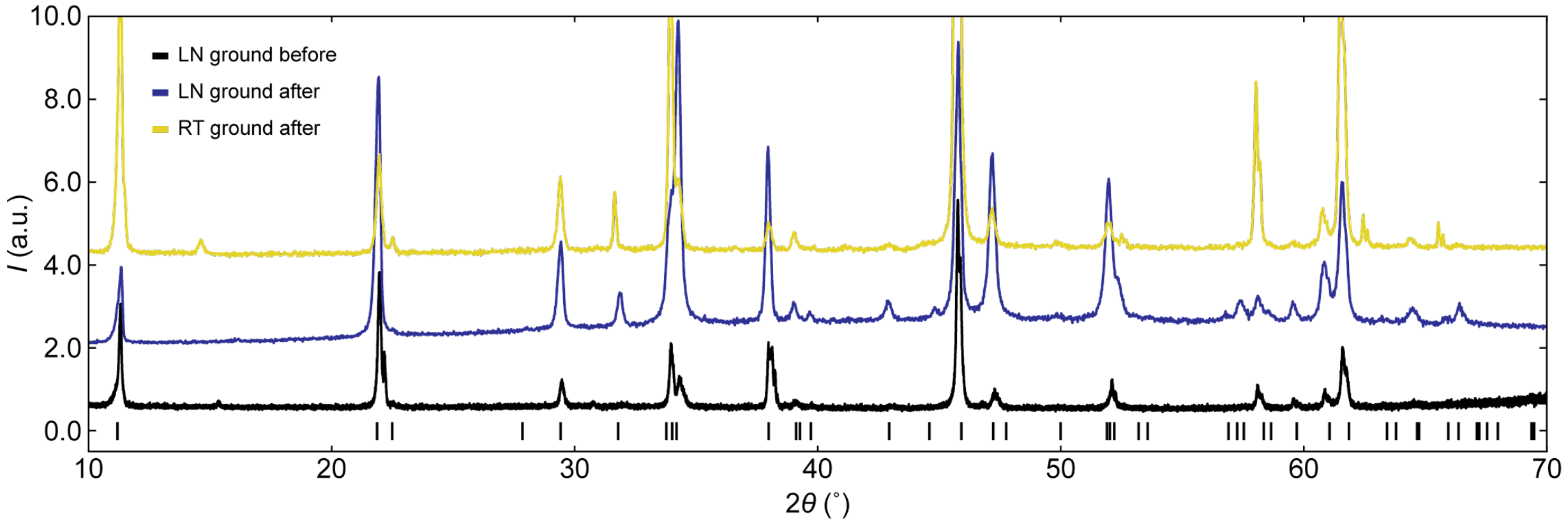

**Figure s10: Powder X-ray diffraction on hydrostatic pressure samples.** Powder X-ray diffraction (PXRD) pattern of the room-temperature-ground (solid yellow trace) and the liquid-nitrogen-ground (solid blue trace) pressure-cell samples after applying maximum hydrostatic pressure. The solid black trace is a reference PXRD pattern of a liquid-nitrogen-ground CrSBr sample before applying pressure. Each pattern is vertically offset for clarity. Short vertical solid black lines are calculated diffraction peak positions.

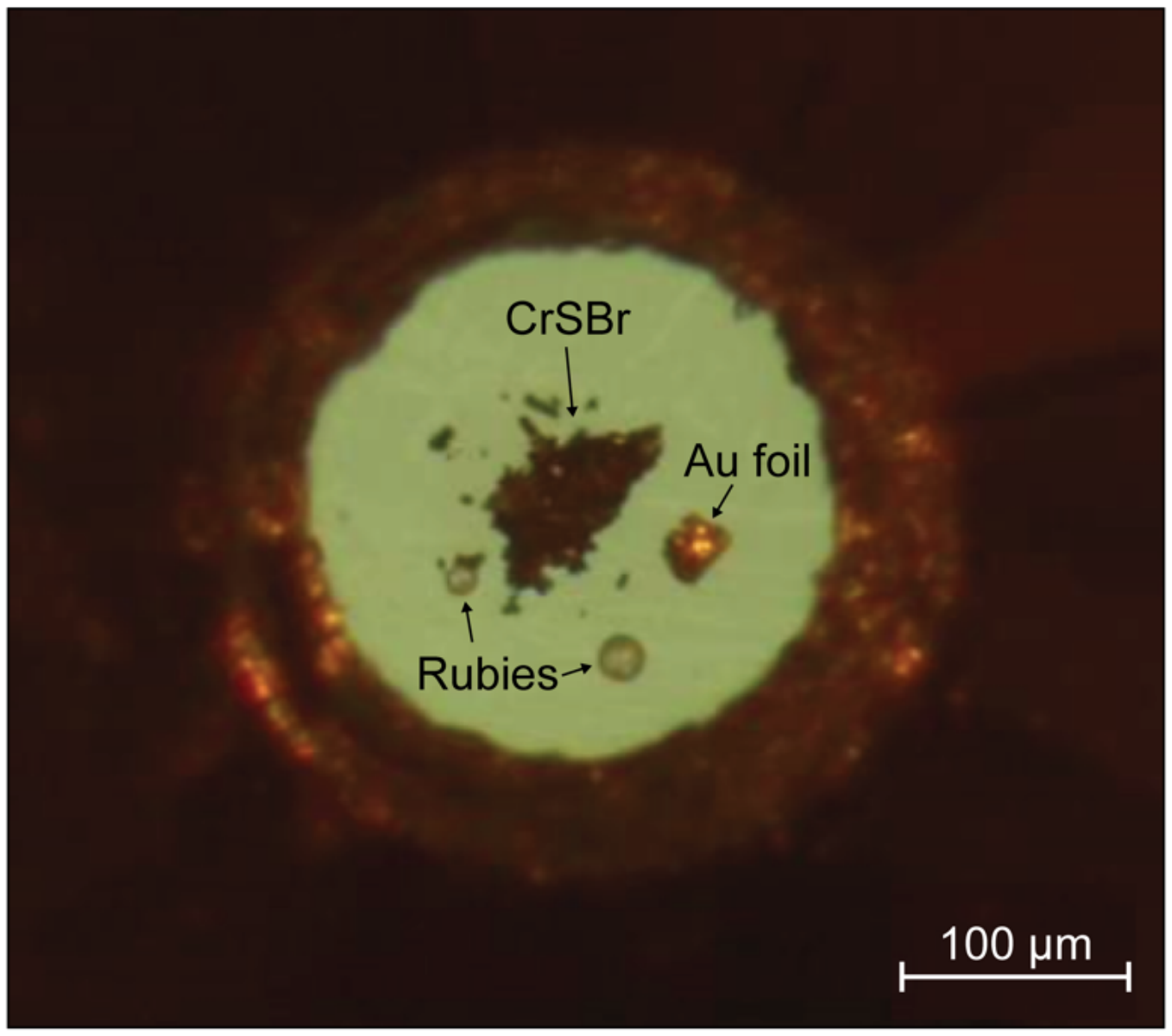

**Figure s11: Photograph of the loaded sample for high pressure diffraction measurements.** Photograph of the sample contained within the diamond anvil cell prior to gas loading. The cell was loaded with CrSBr (center), gold foil (middle right), and two rubies (bottom and bottom left) contained within a stainless-steel gasket. The scale bar is 100 μm.

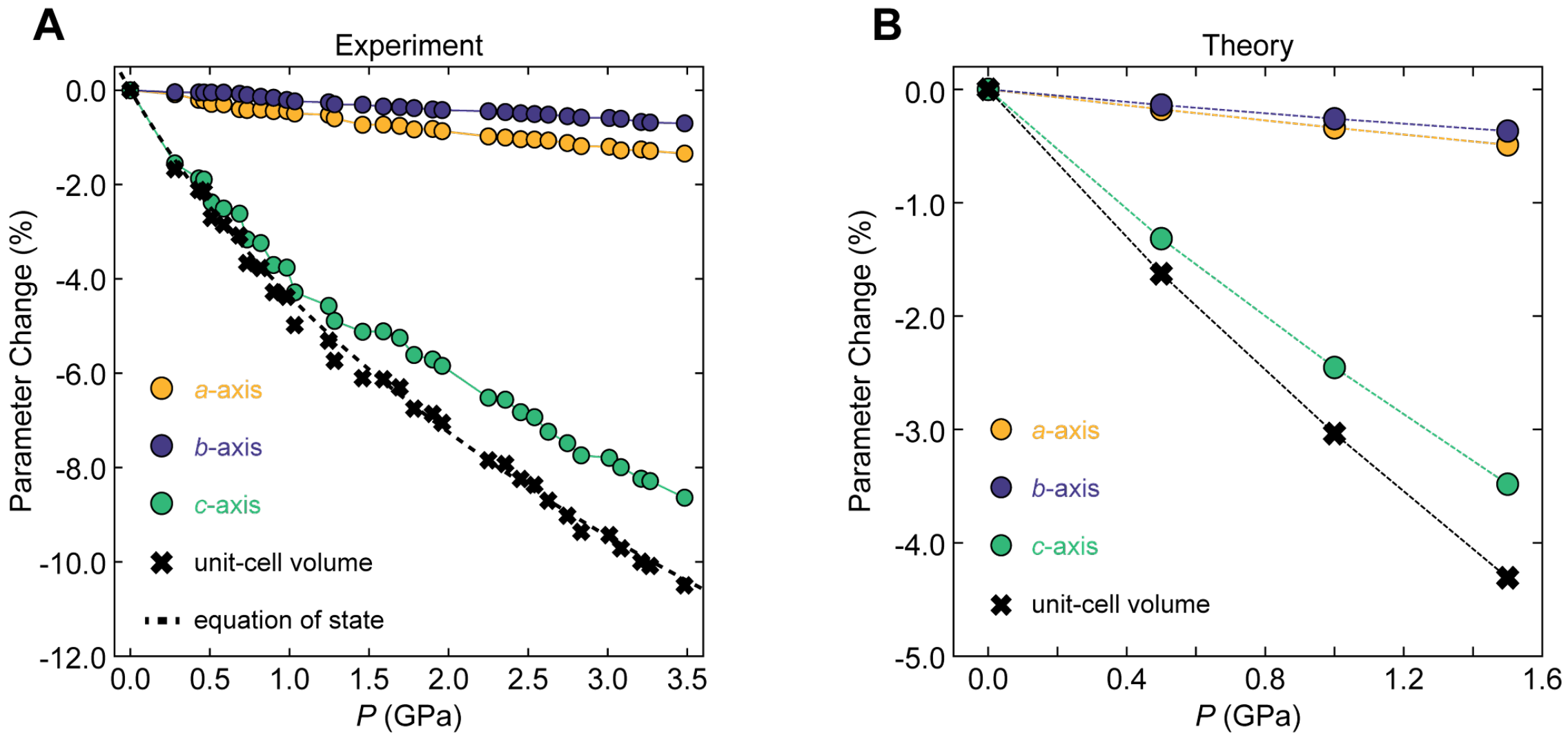

**Figure s12: Comparison between experimental and calculated lattice parameters under pressure. A)** Experimentally-determined percentage change in lattice constants (orange, purple, and green dots for *a*-, *b*-, and *c*-axes, respectively) and the unit-cell volume (black crosses) versus *P*. The dashed black line is a fit to an equation of state. **B)** Calculated percentage change in lattice constants (orange, purple, and green dots for *a*-, *b*-, and *c*-axes, respectively) and the unit-cell volume (black crosses) versus *P*.

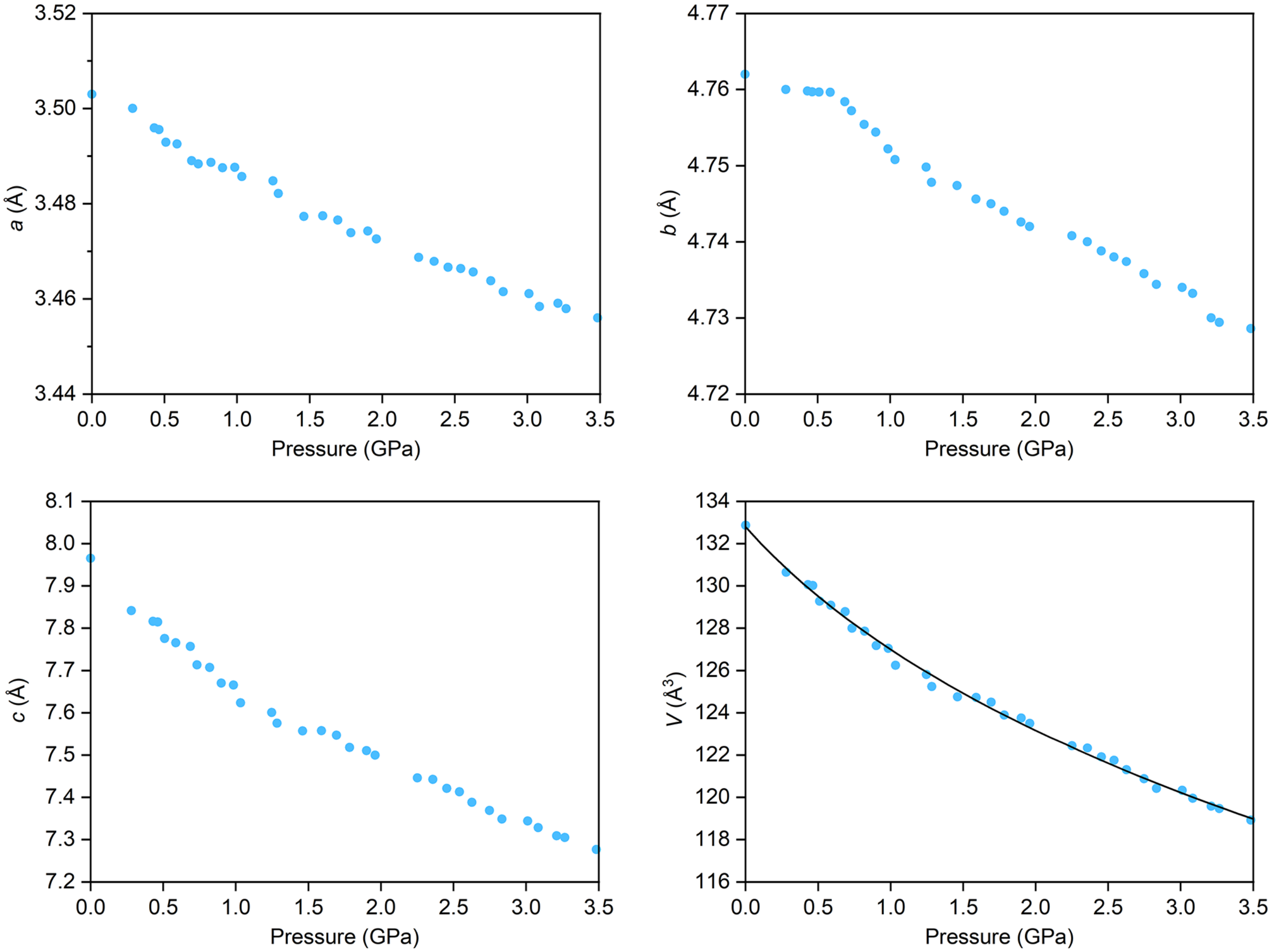

**Figure s13: Pressure-dependent lattice parameters for CrSBr**. Lattice parameters and unit cell volume for CrSBr at each pressure. Lattice parameters were obtained from Pawley refinement of powder X-ray diffraction data. The 0 GPa points were taken from single crystal data at ambient pressure. The line in the unit cell volume plot represents a fit to a third-order Birch-Murnaghan equation of state. The fitted equation of state gave parameters of $V_0$ = 132.785 Å$^3$, $B_0$ = 17.147 GPa, and $B_0$' = 12.626, where $V_0$ is the ambient pressure unit cell volume, $B_0$ is the bulk modulus, and $B_0$' is the first derivative of the bulk modulus with respect to pressure.

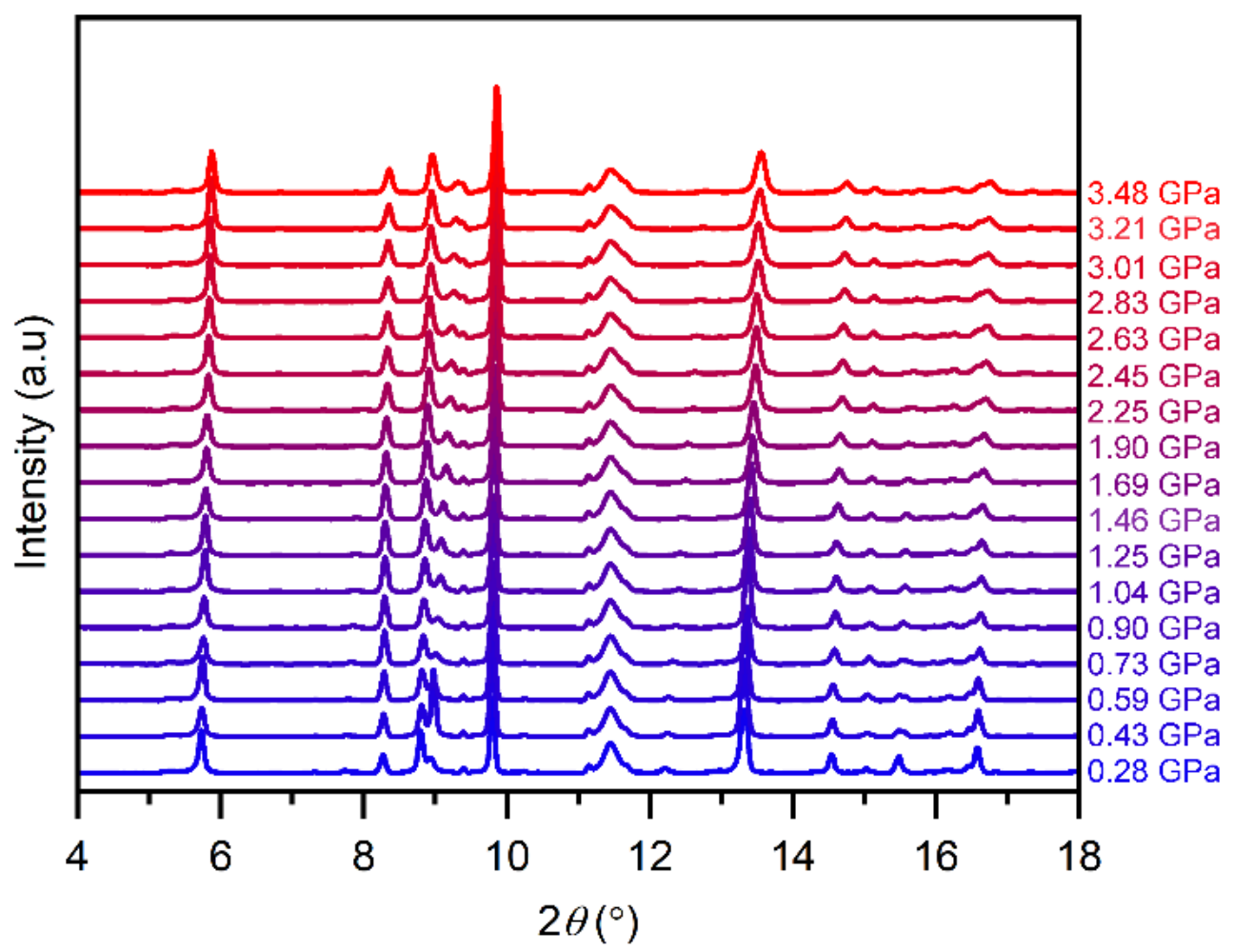

**Figure s14: Pressure-dependent powder X-ray diffraction data.** Normalized, background-corrected powder X-ray diffraction data for CrSBr with an arbitrary vertical offset, collected at a wavelength of 0.406626 Å. Pressures for each pattern are shown to the right.

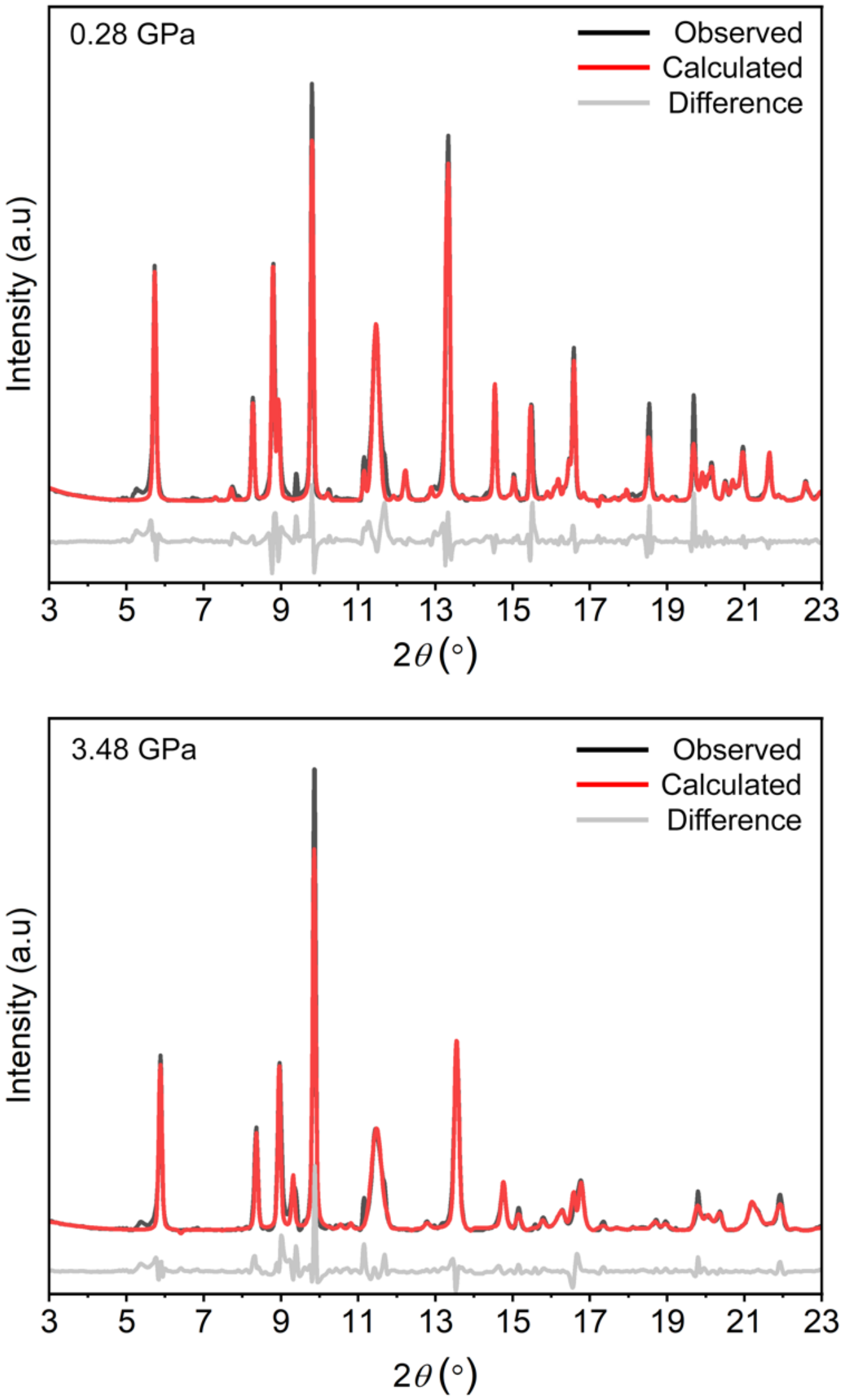

**Figure s15: Representative Pawley refinements for powder X-ray diffraction data.** Powder X-ray diffraction data and Pawley refinements for CrSBr at 0.28 GPa (top) and 3.48 GPa (bottom). Black, red, and gray lines correspond to the observed data, the calculated fit, and the difference, respectively. The broad feature near 11.5° was caused by the sample holder and was modeled as part of the background.

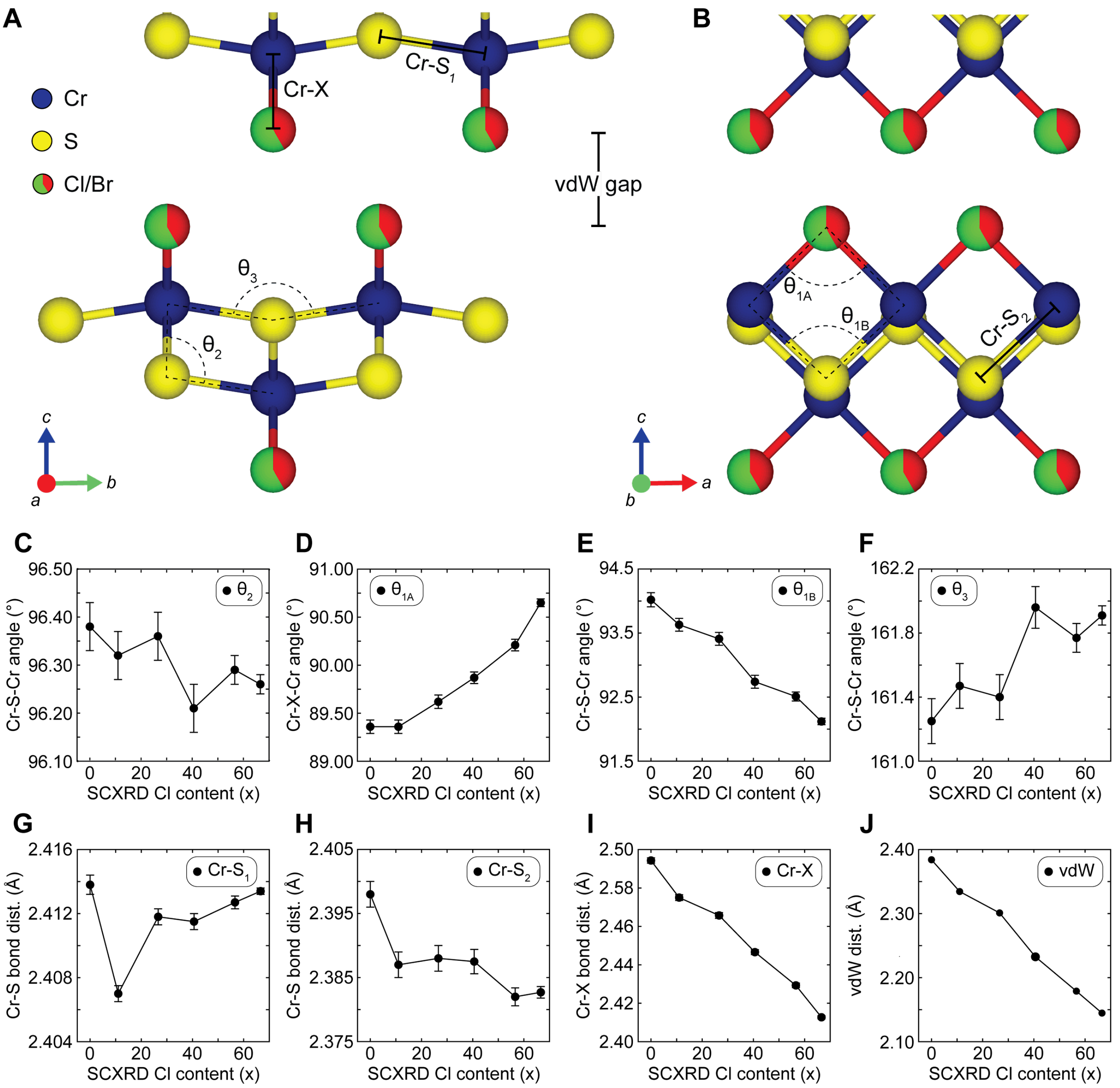

**Figure s16: Selected bond angles and distances for CrSBr and Cl-alloyed CrSBr. A, B)** Crystal structure of Cl-alloyed CrSBr as viewed along the *a*- (**A**) and *b*- (**B**) axis. Select bond angles and distances related to the magnetic couplings $J_1$, $J_2$, $J_3$, and $J_4$ are labeled. **C-F)** Bond angles, **G-I)** bond distances, and **J)** van der Waals gap versus Cl concentration for all growth batches of CrSBr and Cl-alloyed CrSBr.

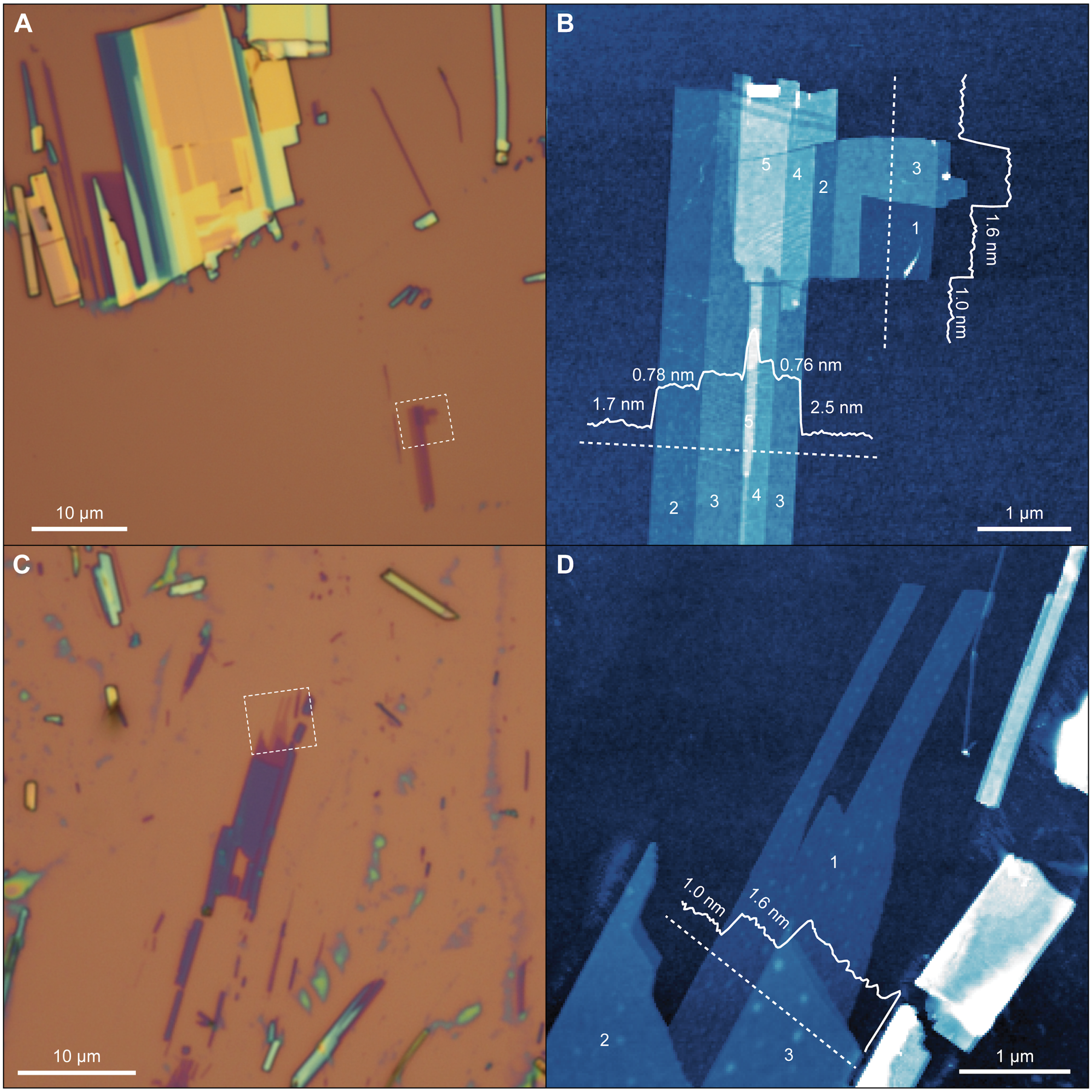

**Figure s17: Exfoliation of $CrSBr_{1-x}Cl_x$. A, C)** Optical microscopy images of $CrSBr_{0.43}Cl_{0.57}$ (**A**) and $CrSBr_{0.33}Cl_{0.67}$ (**C**) mechanically exfoliated onto 285 nm thick $SiO_2$/Si substrates. **B, D)** Atomic force microscopy images of the regions defined by the dashed white boxes in (**A**) and (**C**), respectively. Dashed and solid white lines plot the direction and height profile of select line traces, respectively. Extracted step heights for each line trace are given in the inset. The determined layer numbers for each region of the flakes are denoted. The expected height for 1 layer of $CrSBr_{1-x}Cl_x$ ranges from 0.80 nm down to 0.76 nm between 0% and 67% Cl concentration.

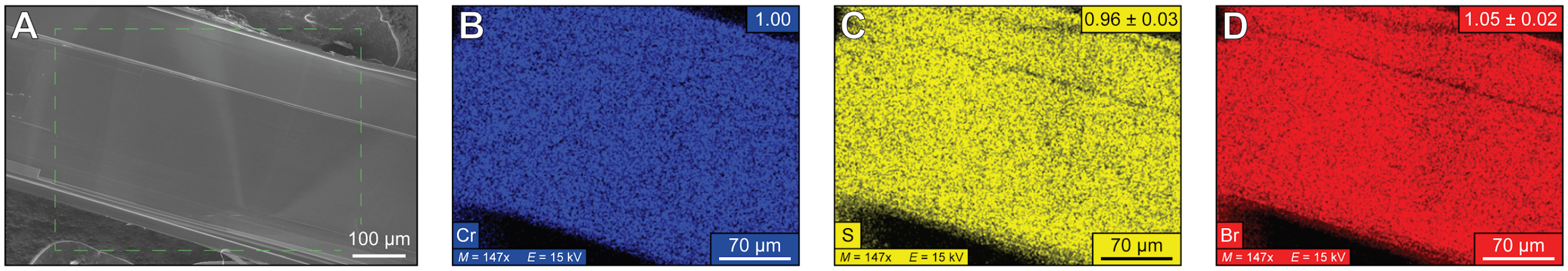

**Figure s18: SEM/EDX mapping of CrSBr. A)** SEM image of a bulk CrSBr crystal. **B-D)** Corresponding EDX chemical maps for Cr (**B**), S (**C**), and Br (**D**) for the region in (**A**) defined by the dashed green box. Extracted chemical percentages are given in the top-right insets. Error bars represent the standard deviation of 7 measurements across 2 crystals.

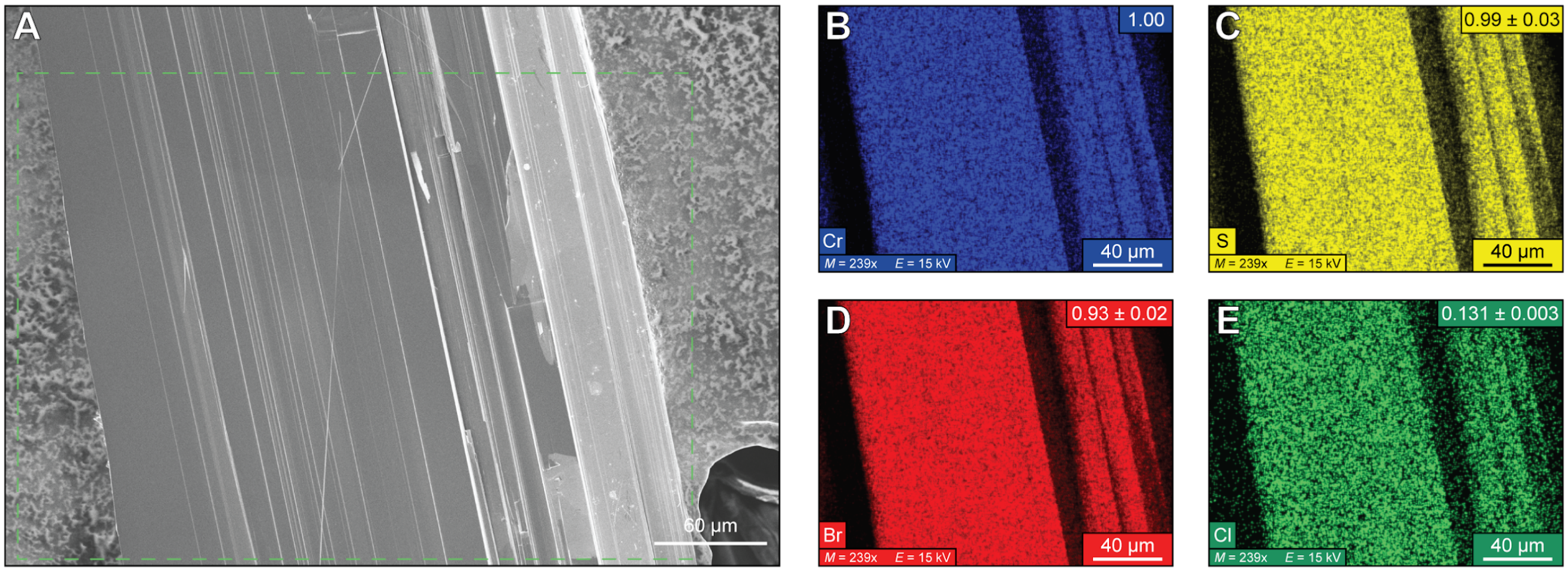

**Figure s19: SEM/EDX mapping of $CrSBr_{1-x}Cl_x$ ($x$=0.11). A)** SEM image of a bulk $CrSBr_{0.89}Cl_{0.11}$ crystal. **B-E)** Corresponding EDX chemical maps for Cr (**B**), S (**C**), Br (**D**), and Cl (**E**) for the region in (**A**) defined by the dashed green box. Extracted chemical percentages are given in the top-right insets. Error bars represent the standard deviation of 10 measurements across 3 crystals.

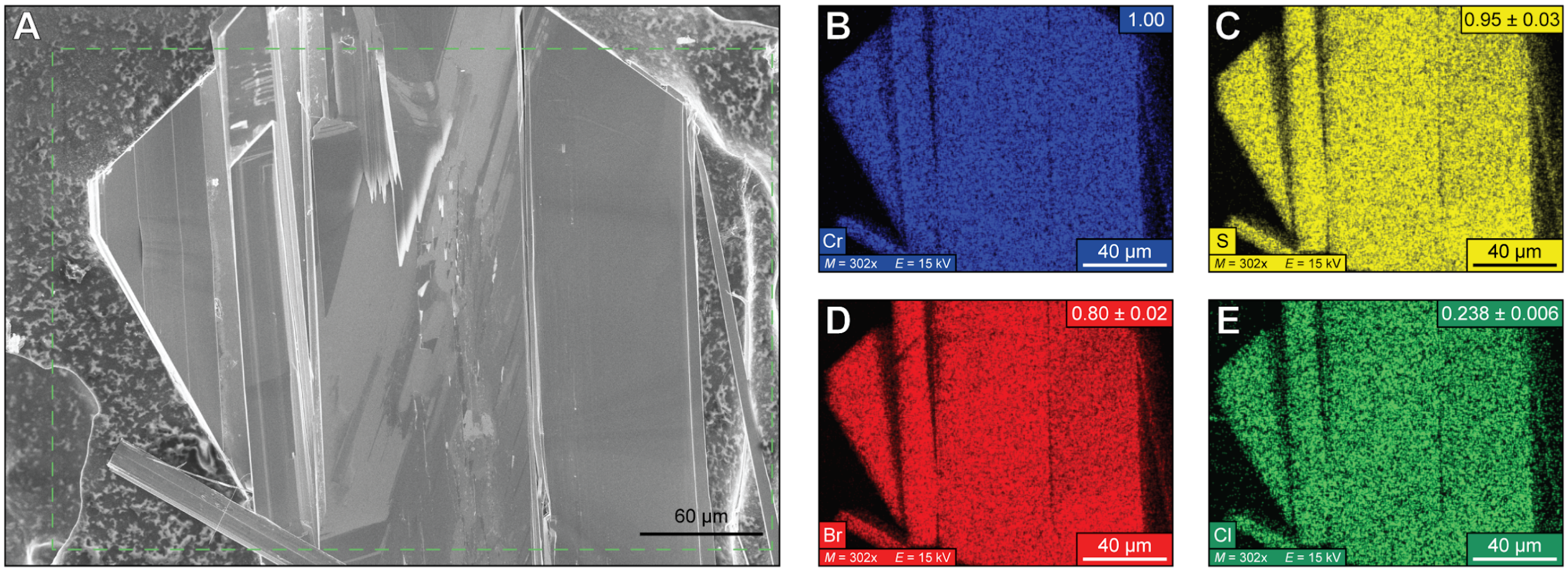

**Figure s20: SEM/EDX mapping of $CrSBr_{1-x}Cl_x$ ($x$=0.27). A)** SEM image of a bulk $CrSBr_{0.73}Cl_{0.27}$ crystal. **B-E)** Corresponding EDX chemical maps for Cr (**B**), S (**C**), Br (**D**), and Cl (**E**) for the region in (**A**) defined by the dashed green box. Extracted chemical percentages are given in the top-right insets. Error bars represent the standard deviation of 9 measurements across 3 crystals.

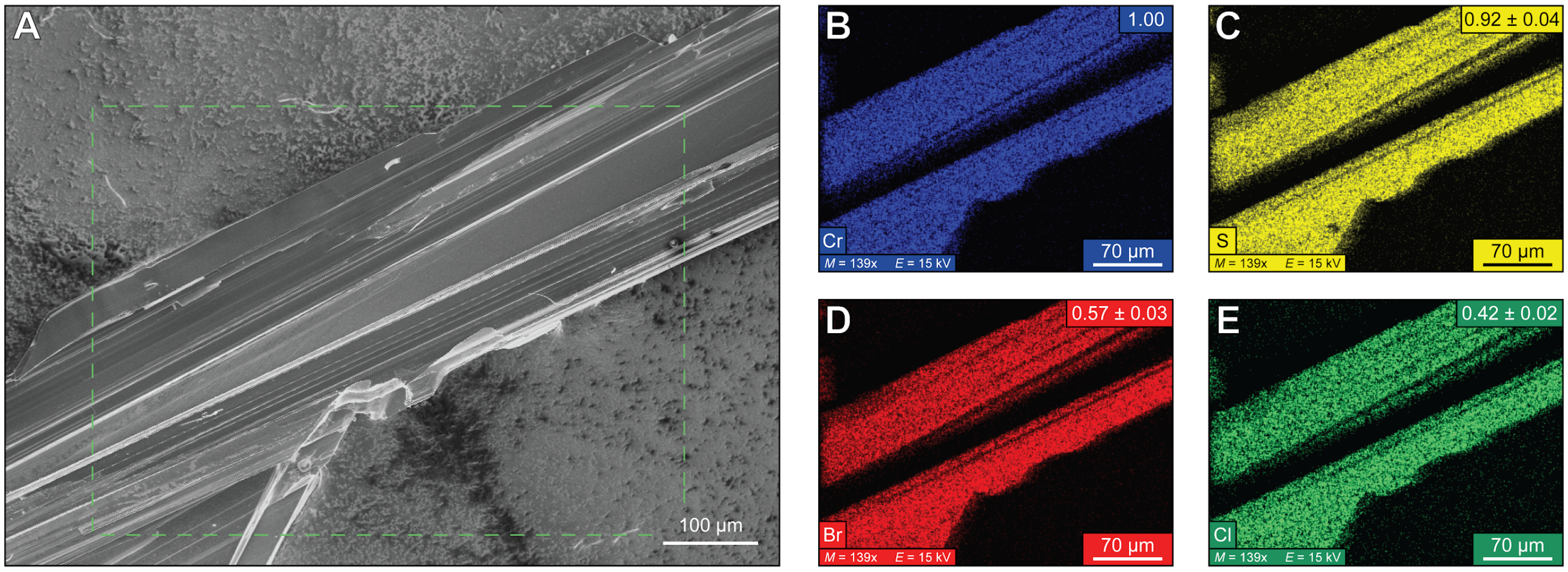

**Figure s21: SEM/EDX mapping of $CrSBr_{1-x}Cl_x$ ($x$=0.41). A)** SEM image of a bulk $CrSBr_{0.59}Cl_{0.41}$ crystal. **B-E)** Corresponding EDX chemical maps for Cr (**B**), S (**C**), Br (**D**), and Cl (**E**) for the region in (**A**) defined by the dashed green box. Extracted chemical percentages are given in the top-right insets. Error bars represent the standard deviation of 13 measurements across 5 crystals.

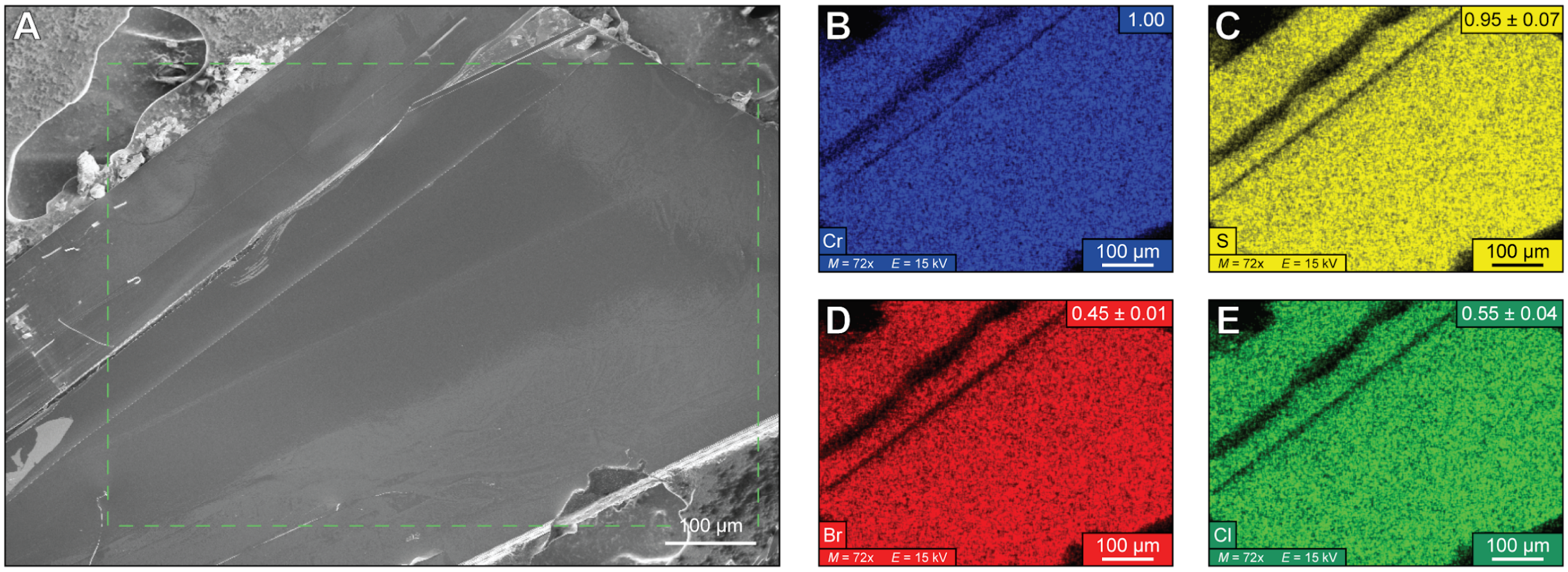

**Figure s22: SEM/EDX mapping of $CrSBr_{1-x}Cl_x$ ($x$=0.57). A)** SEM image of a bulk $CrSBr_{0.43}Cl_{0.57}$ crystal. **B-E)** Corresponding EDX chemical maps for Cr (**B**), S (**C**), Br (**D**), and Cl (**E**) for the region in (**A**) defined by the dashed green box. Extracted chemical percentages are given in the top-right insets. Error bars represent the standard deviation of 14 measurements across 6 crystals.

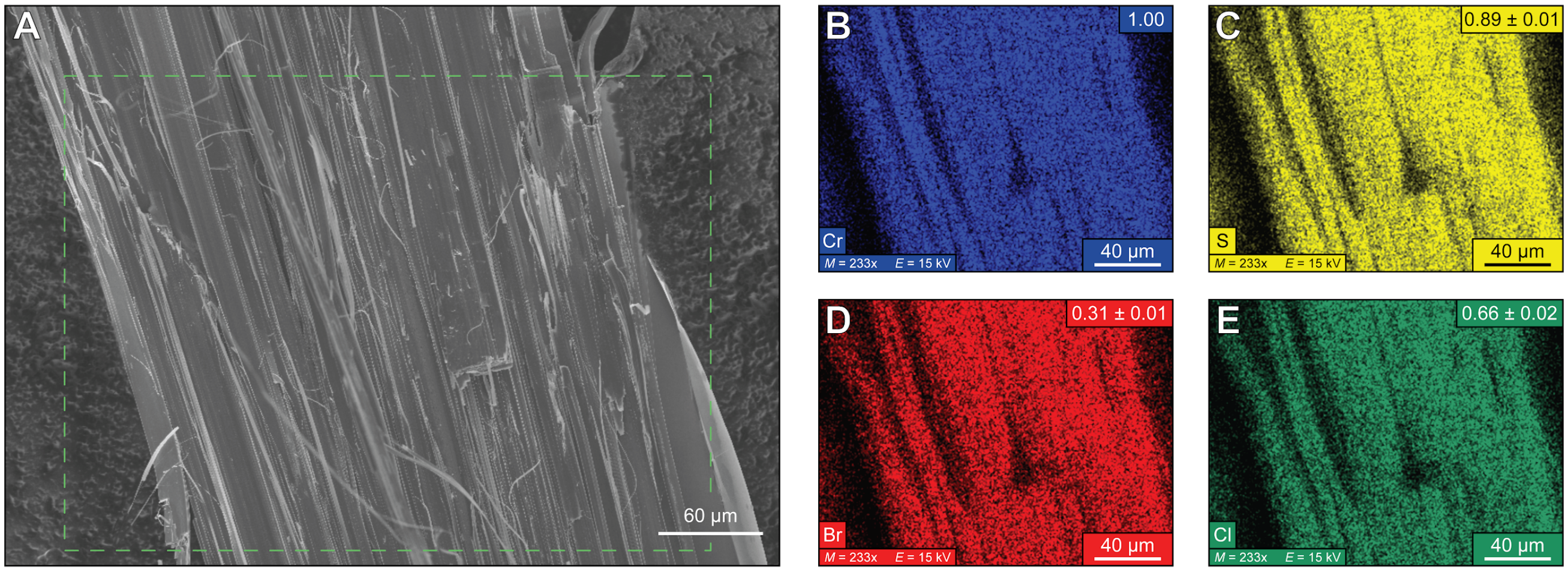

**Figure s23: SEM/EDX mapping of $CrSBr_{1-x}Cl_x$ ($x$=0.67). A)** SEM image of a bulk $CrSBr_{0.33}Cl_{0.67}$ crystal. **B-E)** Corresponding EDX chemical maps for Cr (**B**), S (**C**), Br (**D**), and Cl (**E**) for the region in (**A**) defined by the dashed green box. Extracted chemical percentages are given in the top-right insets. Error bars represent the standard deviation of 9 measurements across 3 crystals.

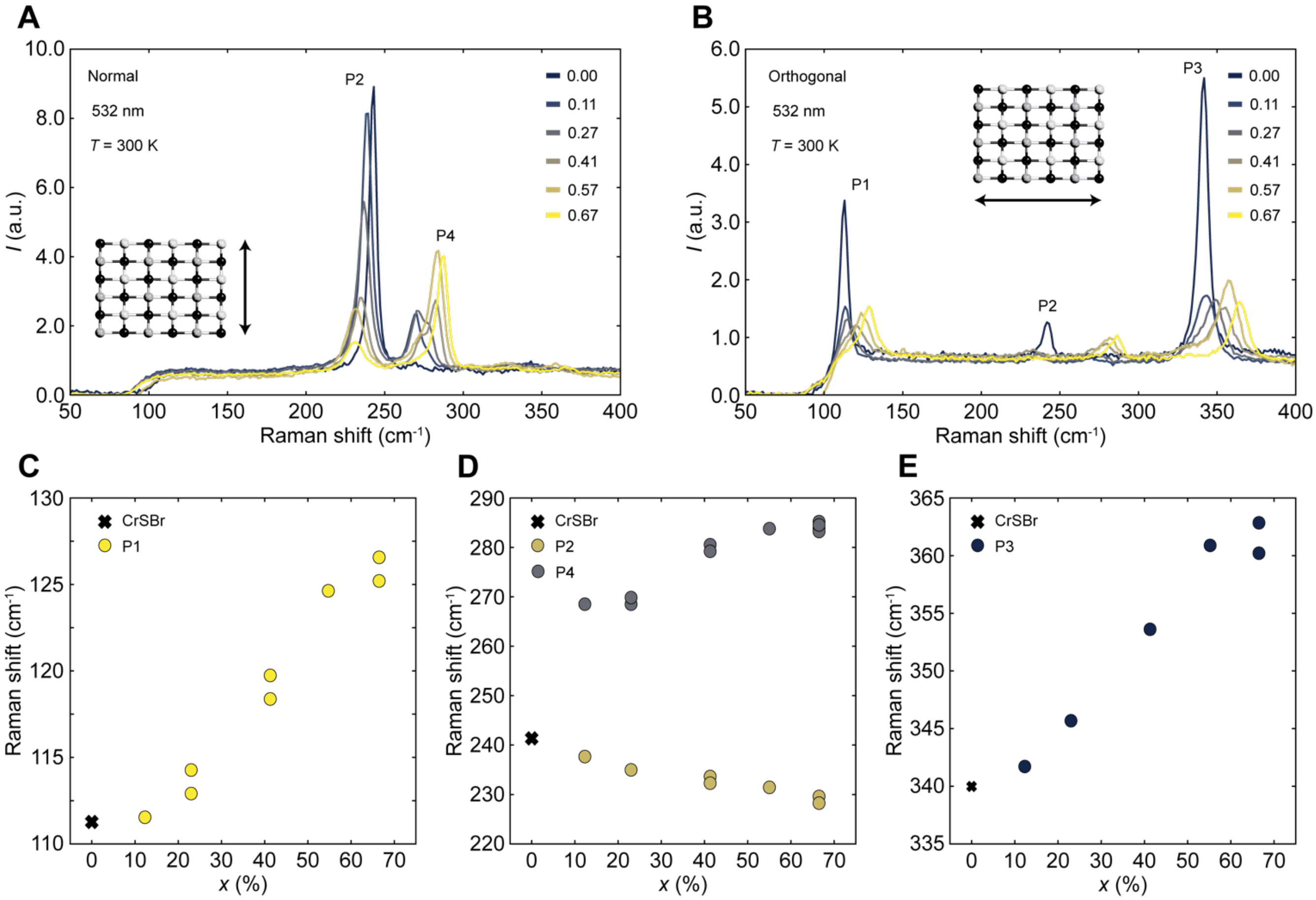

**Figure s24: Raman spectroscopy on bulk $CrSBr_{1-x}Cl_x$. A, B)** Intensity versus Raman shift for bulk $CrSBr_{1-x}Cl_x$ with incident light polarized parallel to the *a*-axis (**A**) and the *b*-axis (**B**) for various Cl doping levels. The corresponding Raman peaks are labeled. **C-E**) Extracted peak positions versus Cl doping for P1 (**C**), P2 and P4 (**D**), and P3 (**E**).

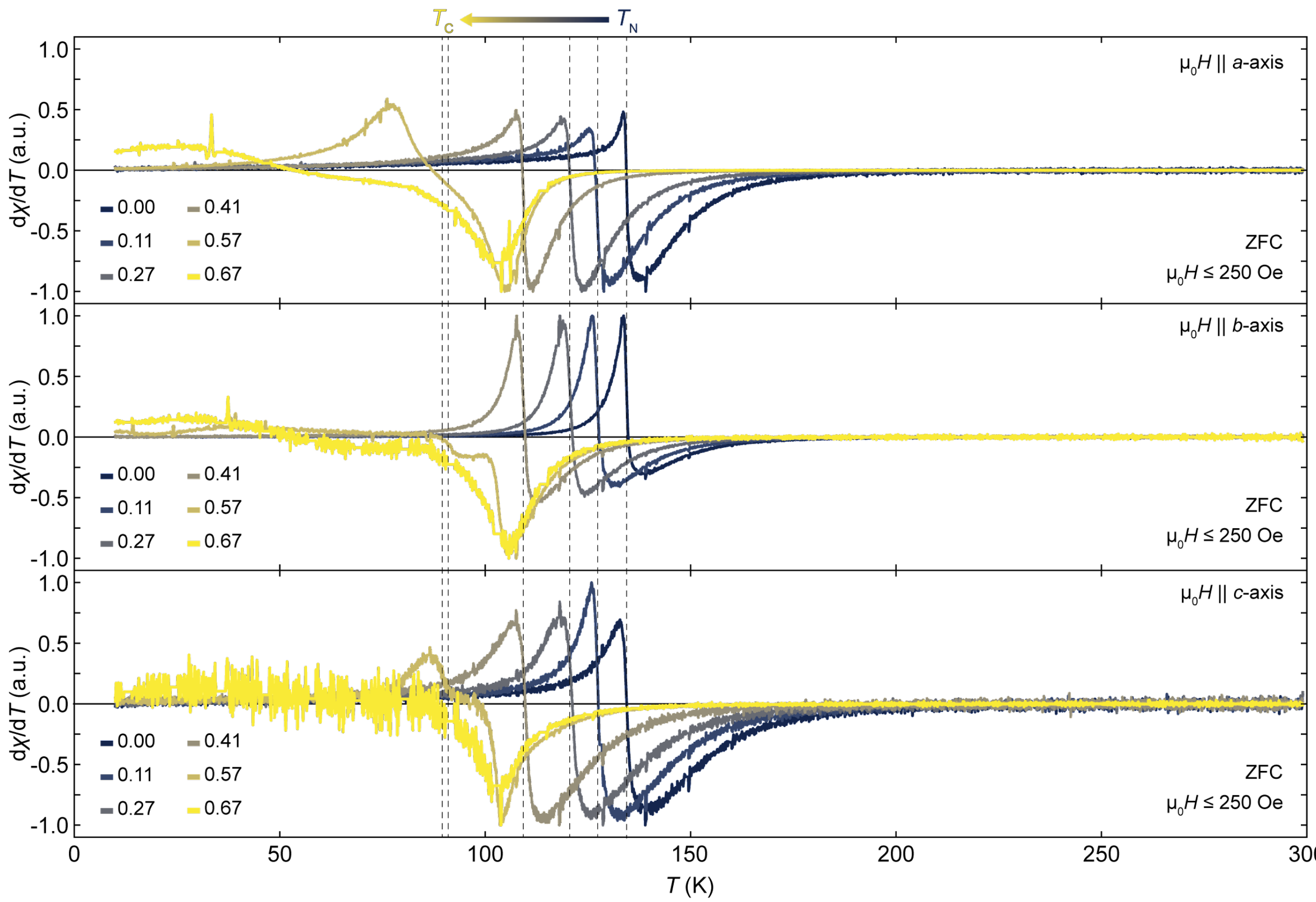

**Figure s25: Determination of $T_N$ for $CrSBr_{1-x}Cl_x$.** Normalized derivative of the zero-field-cooled magnetic susceptibility versus temperature for magnetic fields aligned along the *a*- (top panel), *b*- (middle panel), and *c*-axes (bottom panel) for various Cl doping. The corresponding Cl doping is given in the inset. A measuring field of 250 Oe was used for Cl-00, Cl-11, and Cl-27, whereas a measuring field of 100 Oe was used for Cl-41, Cl-57, and Cl-67. The clear zero-crossing in the Cl-00, Cl-11, Cl-27, and Cl-41 traces denotes $T_N$. In the Cl-57 and Cl-67 traces, $T_C$ is defined using the ac susceptibility measurements in **figure s28** and **figure s29**, respectively. The extracted critical temperature for each Cl doping is denoted by black dashed lines.

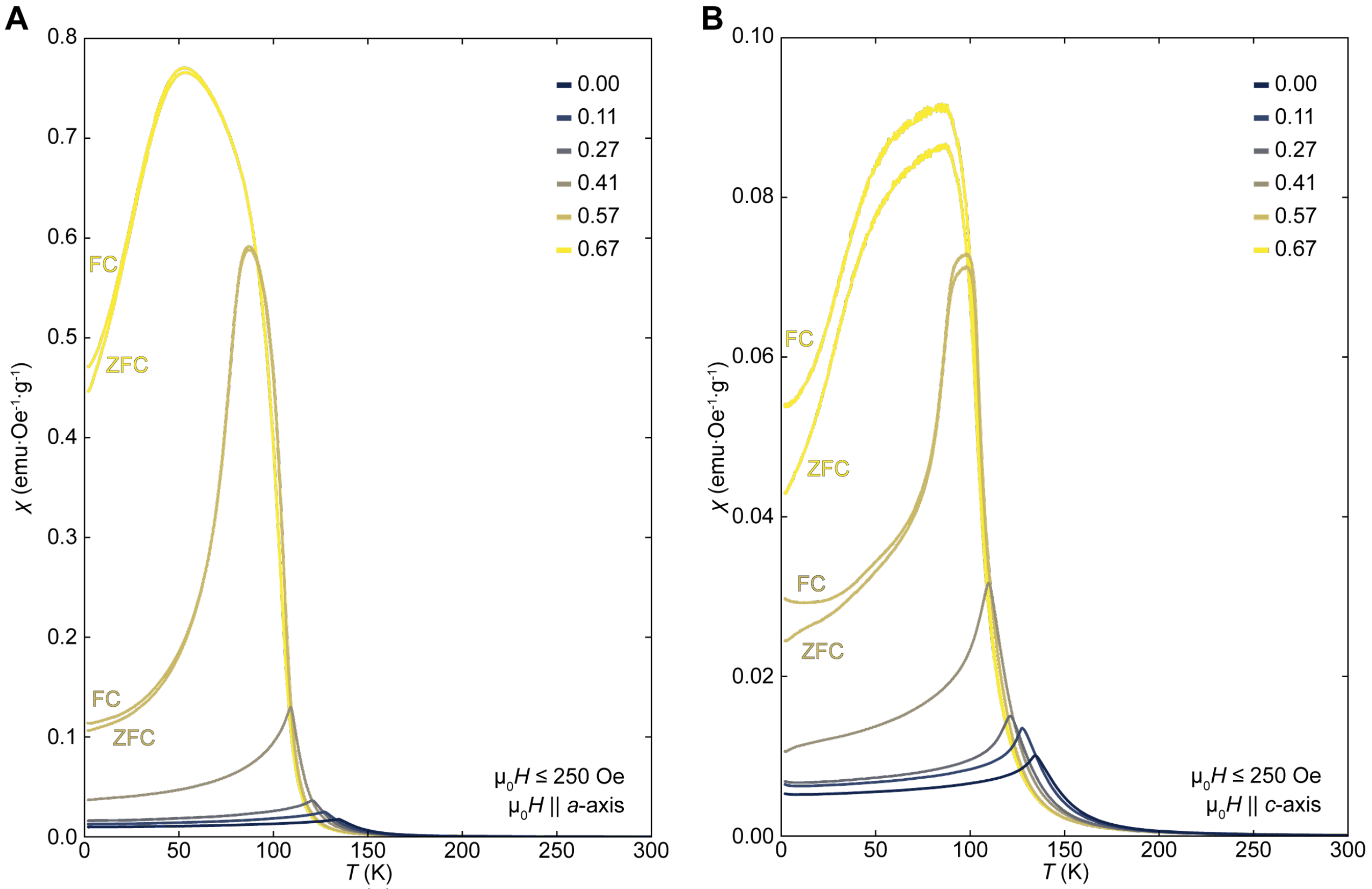

**Figure s26: Additional magnetic susceptibility data on $CrSBr_{1-x}Cl_x$. A)** Magnetic susceptibility versus temperature with the magnetic field aligned along the *a*-axis for various levels of Cl doping. The corresponding Cl doping is given in the inset. A measuring field of 250 Oe was used for Cl-00, Cl-11, and Cl-27, whereas a measuring field of 100 Oe was used for Cl-41, Cl-57, and Cl-67. For Cl-00, Cl-11, Cl-27, and Cl-41, only the zero-field trace is shown as it overlaps the field-cooled trace. For Cl-57 and Cl-67, both zero-field-cooled and field-cooled traces are shown. **B)** Magnetic susceptibility versus temperature with the magnetic field aligned along the *c*-axis for various levels of Cl doping. The corresponding Cl doping is given in the inset. A measuring field of 250 Oe was used for Cl-00, Cl-11, and Cl-27, whereas a measuring field of 100 Oe was used for Cl-41, Cl-57, and Cl-67. For Cl-00, Cl-11, Cl-27, and Cl-41, only the zero-field trace is shown as it overlaps the field-cooled trace. For Cl-57 and Cl-67, both zero-field-cooled and field-cooled traces are shown.

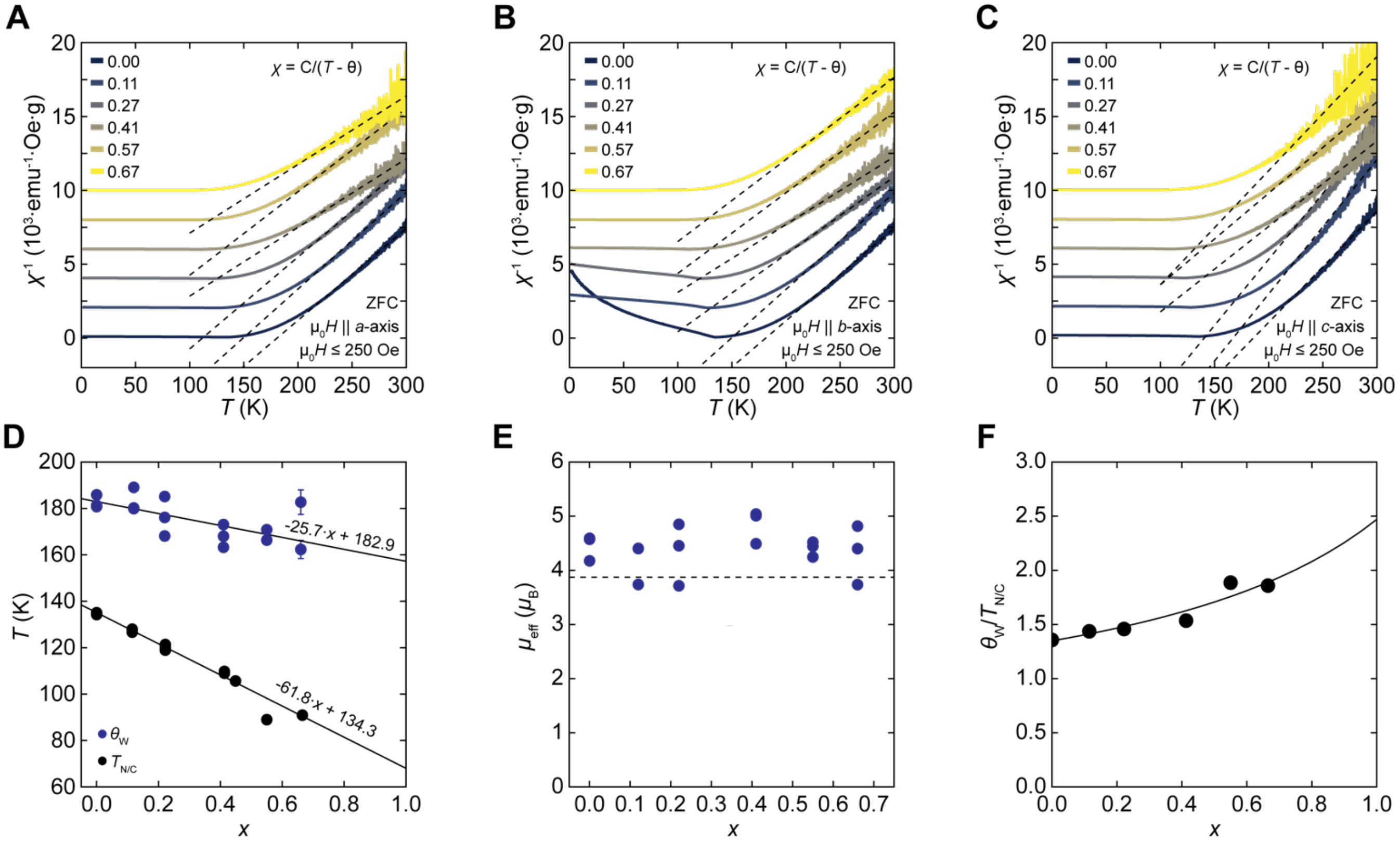

**Figure s27: Curie-Weiss analysis of $CrSBr_{1-x}Cl_x$. A-C)** Inverse susceptibility versus temperature of $CrSBr_{1-x}Cl_x$ for magnetic fields oriented along the *a*- (**A**), *b*- (**B**), and *c*-axis (**C**) for various Cl doping. Dashed black lines correspond to linear Curie-Weiss fits. **D)** Extracted Weiss temperature (blue dots) along with the extracted critical temperature (black dots) versus Cl doping. Solid black lines are linear fits to the data. The resulting fit parameters are given in the inset. **E)** Extracted effective $Cr^{3+}$ magnetic moment versus Cl doping. The dashed black line is the expected effective magnetic moment for $S = 3/2$ $Cr^{3+}$. **F)** Ratio of Weiss temperature to critical temperature versus Cl doping. The solid black line is the ratio of the linear fit of the Weiss temperature to the linear fit of the critical temperature versus Cl doping.

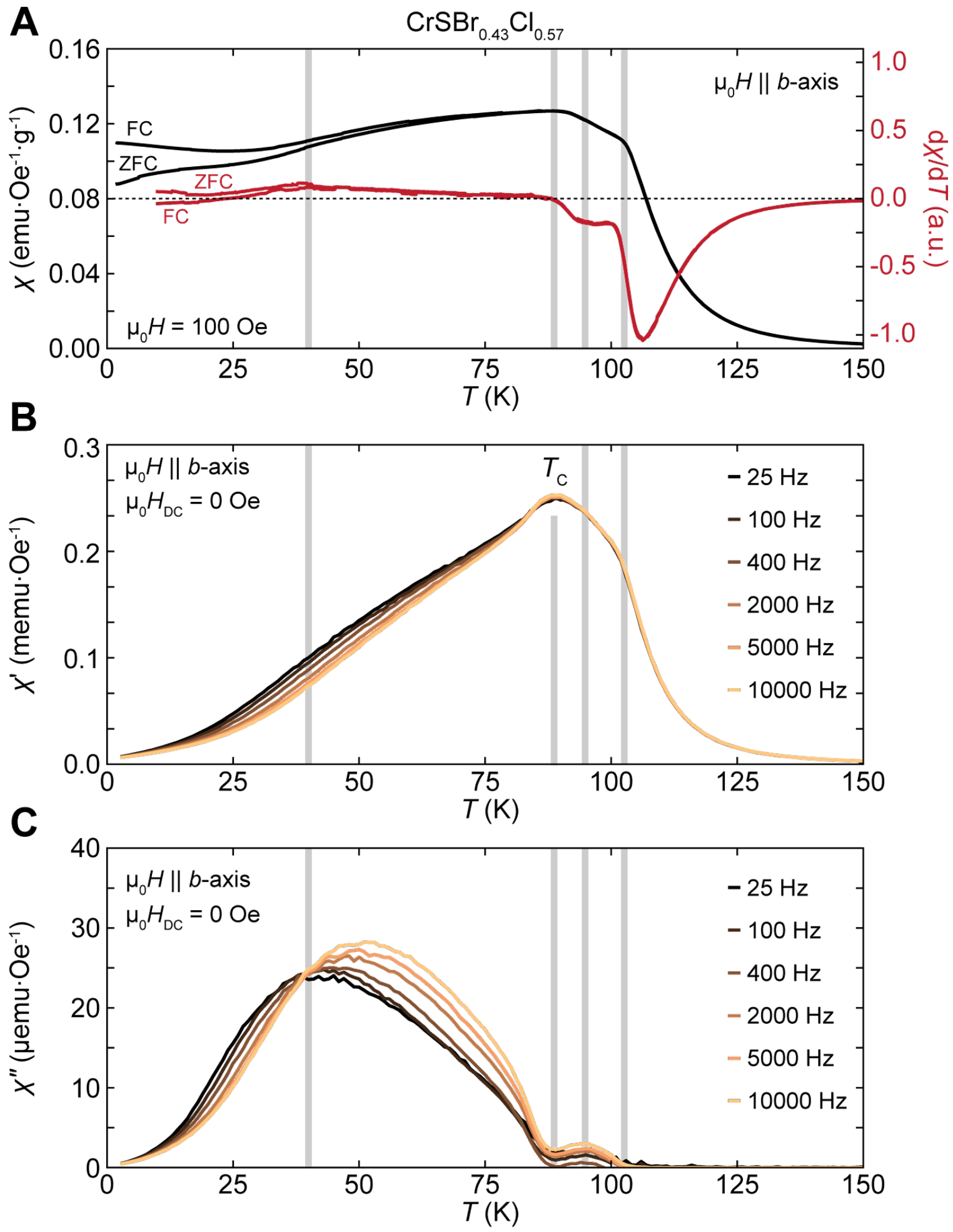

**Figure s28: Ac magnetometry on $CrSBr_{0.43}Cl_{0.57}$. A)** Dc magnetic susceptibility (solid black lines) and the derivative of magnetic susceptibility (solid red lines) versus temperature for magnetic fields aligned parallel to the *b*-axis. A measuring field of 100 Oe was used. Both field-cooled and zero-field-cooled traces are shown. **B, C)** Real (**B**) and imaginary (**C**) components of the ac magnetic susceptibility versus temperature for various ac field frequencies (denoted by different trace colors or line styles). The ac magnetic field was applied parallel to the *b*-axis. Zero ac magnetic field was applied. In all plots, vertical solid gray bars denote distinct features in χ, χ', or χ''.

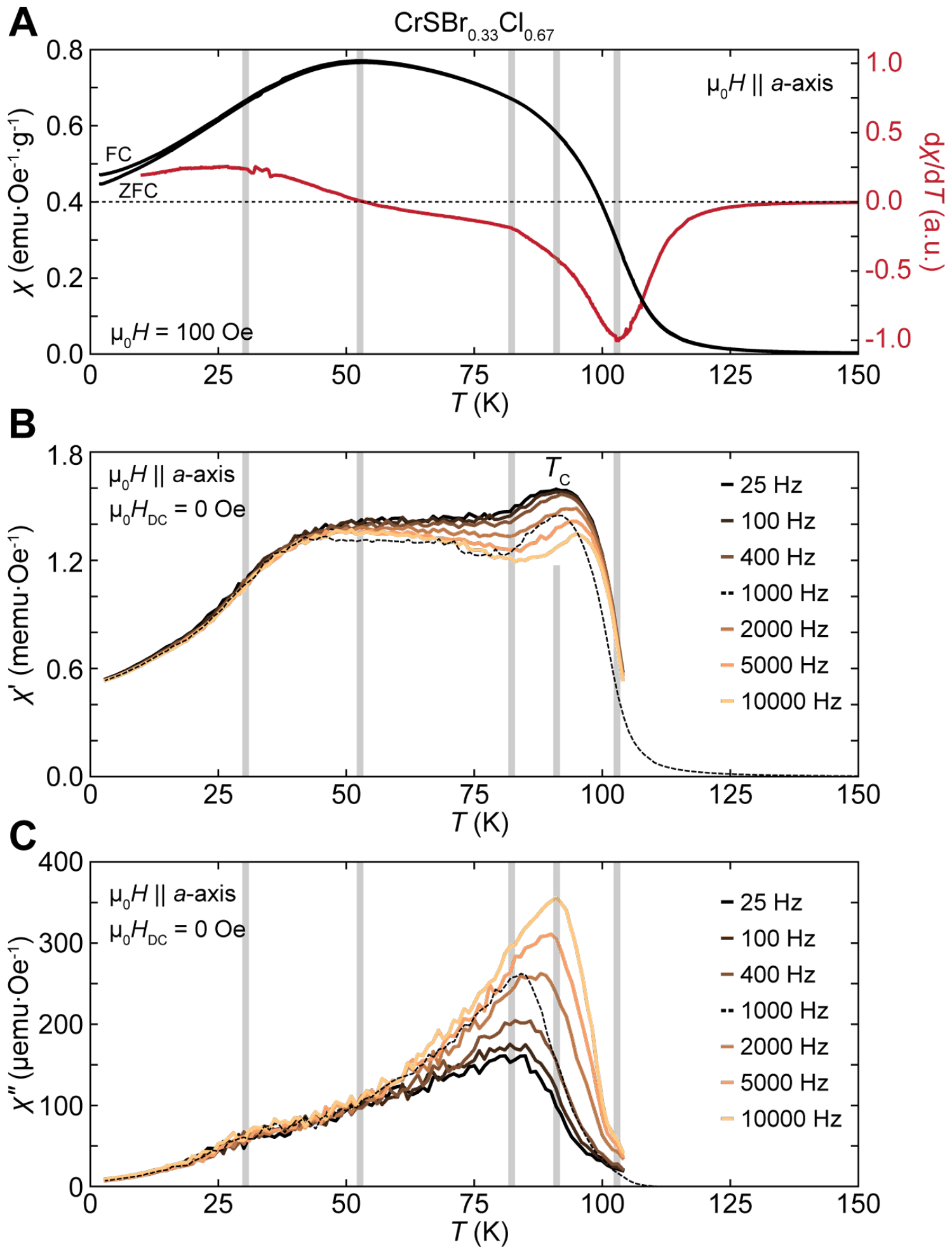

**Figure s29: Ac magnetometry on $CrSBr_{0.33}Cl_{0.67}$. A)** Dc magnetic susceptibility (solid black lines) and the derivative of magnetic susceptibility (solid red lines) versus temperature for magnetic fields aligned parallel to the *a*-axis. A measuring field of 100 Oe was used. Both field-cooled and zero-field-cooled traces are shown. **B, C)** Real (**B**) and imaginary (**C**) components of the ac magnetic susceptibility versus temperature for various ac field frequencies (denoted by different trace colors or line styles). The ac magnetic field was applied parallel to the *a*-axis. Zero dc magnetic field was applied. In all plots, vertical solid gray bars denote distinct features in χ, χ', or χ''.

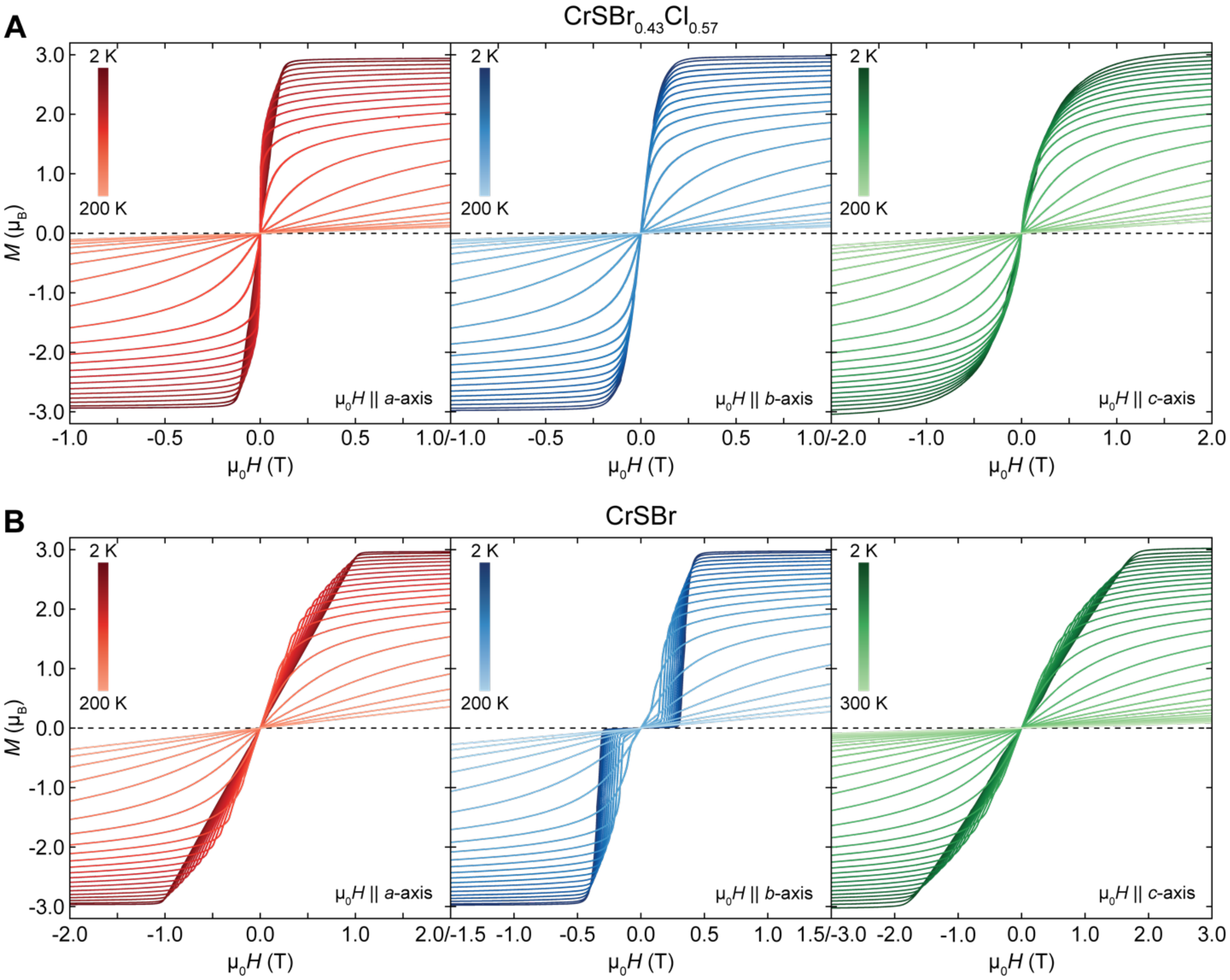

**Figure s30: Comparison of magnetic properties between CrSBr and $CrSBr_{0.43}Cl_{0.57}$. A)** Magnetization versus magnetic field for $CrSBr_{0.43}Cl_{0.57}$ at various temperatures for magnetic fields oriented along the *a*- (red traces), *b*- (blue traces), and *c*-axes (green traces). **B)**. Magnetization versus magnetic field for CrSBr at various temperatures for magnetic fields oriented along the *a*- (red traces), *b*- (blue traces), and *c*-axes (green traces). For each panel, the temperature ranges over which the magnetization traces were acquired is given in the inset.

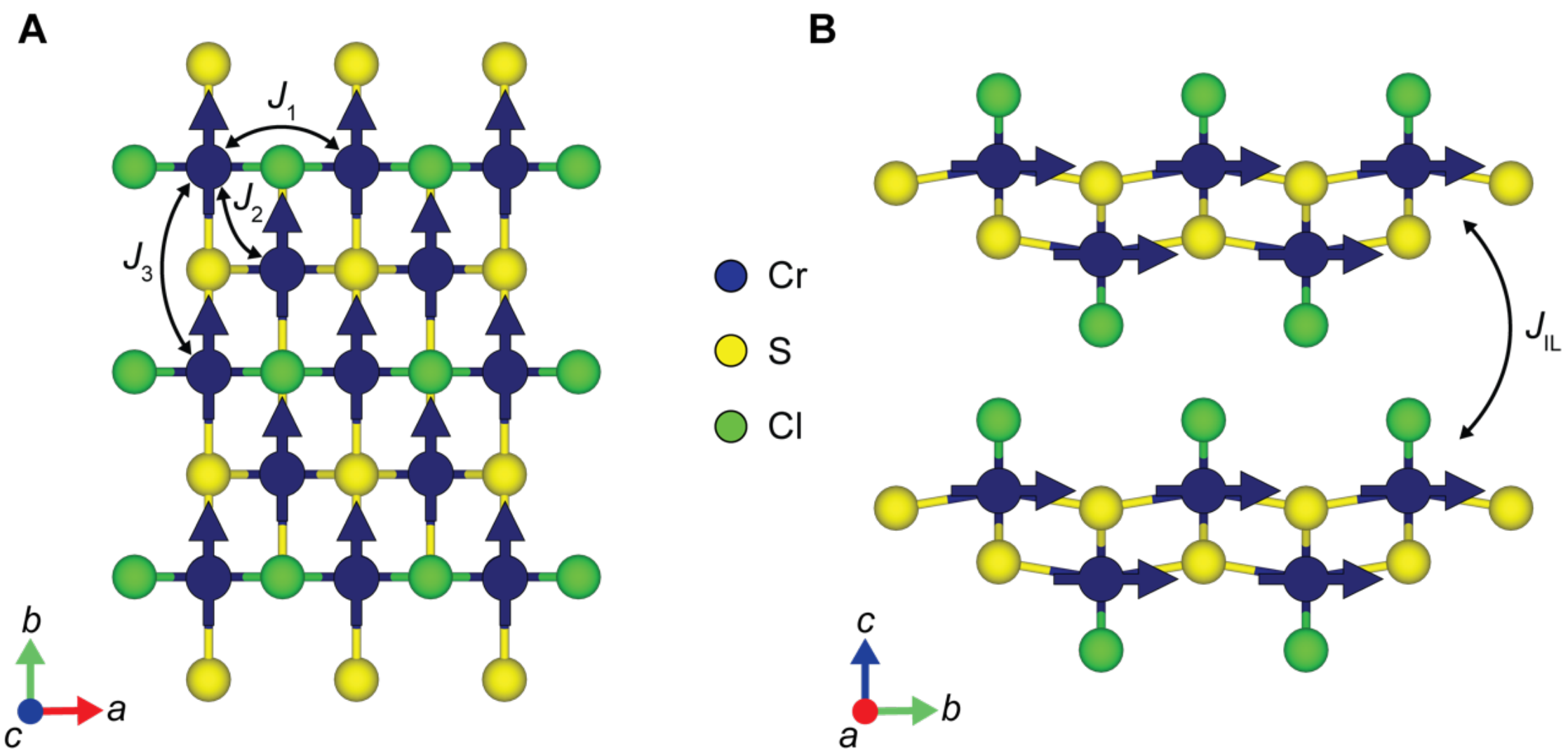

**Figure s31: Calculated crystal structure and magnetic ground state of CrSCl. A, B)** Calculated crystal structure of CrSCl as viewed along the *c*-axis (**A**) and the *a*-axis (**B**). Orientation of the Cr spins in the calculated ferromagnetic ground state are given by solid blue arrows. Blue, yellow, and green circles correspond to Cr, S, and Cl, respectively.

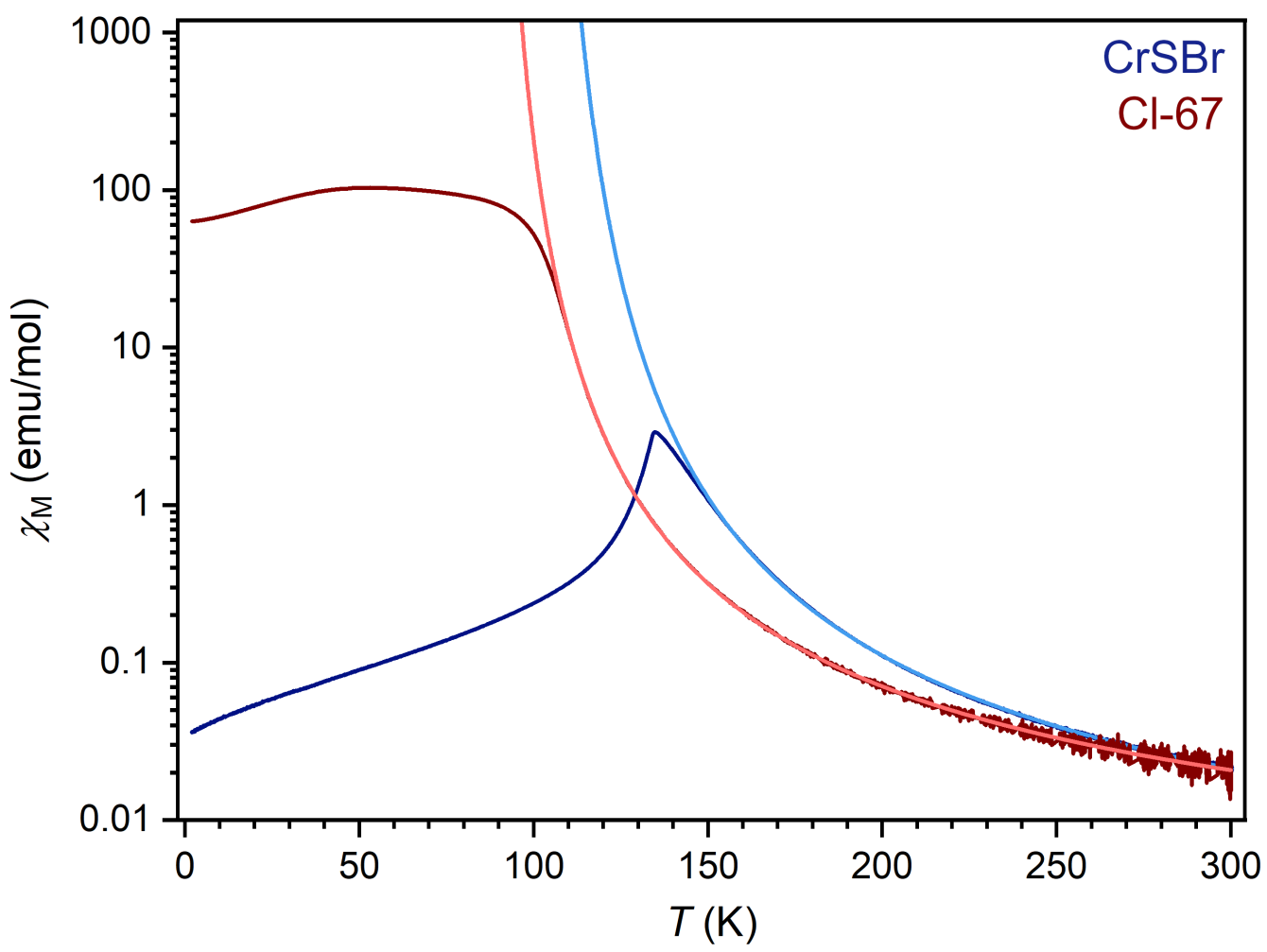

**Figure s32: Fit of the CrSBr and $CrSBr_{0.33}Cl_{0.67}$ susceptibility to the Kosterlitz-Thouless model.** Fits (light blue and light red lines) of the high-temperature susceptibility (dark blue and dark red lines) of CrSBr and Cl-67 to the Kosterlitz-Thouless model: $\chi(T) = A \cdot e^{B \cdot t^{-\frac{1}{2}}}$, where $A$ and $B$ are fit parameters and $t$ is the reduced temperature $t = (T\text{-}T_{KT})/T_{KT}$. Here, $T_{KT}$ is also a fit parameter and represents the temperature below which magnetic vortices should bind. For CrSBr and Cl-67, we obtain fitted $T_{KT}$ of 99 K and 84 K, respectively. While the ordering temperature of CrSBr is much larger than the $T_{KT}$ predicted by its high temperature susceptibility, the freezing temperature of Cl-67 is comparable to its predicted $T_{KT}$, consistent with its weaker interlayer interactions and suppressed anisotropy energy between the *a*- and *b*-axes. For both materials, the strong agreement between the Kosterlitz-Thouless fits and the magnetic susceptibility above the Weiss temperature suggest the possibility of XY-like behavior within the paramagnetic phase.

| | $P$ (GPa) | 0.0 | 0.5 | 1.0 | 1.5 |
|---|---|---|---|---|---|
| Structural Properties | a (Å) | 3.516 | 3.510 | 3.504 | 3.499 |
| | b (Å) | 4.718 | 4.711 | 4.705 | 4.700 |
| | c (Å) | 8.031 | 7.907 | 7.816 | 7.734 |
| | V (Å$^3$) | 132.909 | 130.747 | 128.873 | 127.182 |
| | $\Theta_2$ (°) | 96.505 | 96.416 | 96.321 | 96.222 |
| | $\Theta_{1A}$ (°) | 88.801 | 88.825 | 88.862 | 88.905 |
| | $\Theta_{1B}$ (°) | 95.210 | 95.105 | 95.014 | 94.939 |
| | $\Theta_3$ (°) | 160.657 | 160.940 | 161.240 | 161.551 |
| | Cr-$S_1$ (Å) | 2.393 | 2.389 | 2.385 | 2.381 |
| | Cr-$S_2$ (Å) | 2.380 | 2.378 | 2.376 | 2.374 |
| | Cr-Br (Å) | 2.513 | 2.508 | 2.503 | 2.498 |
| | vdW gap (Å) | 2.416 | 2.324 | 2.249 | 2.181 |
| | $J_1$ Cr-Cr (Å) | 3.516 | 3.510 | 3.504 | 3.499 |
| | $J_2$ Cr-Cr (Å) | 3.561 | 3.554 | 3.547 | 3.540 |
| | $J_3$ Cr-Cr (Å) | 4.718 | 4.711 | 4.705 | 4.700 |
| Magnetic Properties | $J_1$ (meV) | -3.165 | -2.906 | -2.665 | -2.443 |
| | $J_2$ (meV) | -6.495 | -6.425 | -6.317 | -6.241 |
| | $J_3$ (meV) | -5.702 | -5.475 | -5.234 | -4.978 |
| | $J_{IL}$ (meV) | 0.044 | 0.090 | 0.140 | 0.194 |
| | $H_{sat\text{-}b}$ (T) | 0.039 | 0.306 | 0.608 | 0.943 |
| | $H_{sat\text{-}a}$ (T) | 0.121 | 0.384 | 0.680 | 1.008 |
| | $H_{sat\text{-}c}$ (T) | 0.603 | 0.879 | 1.189 | 1.531 |
| | $T_N$ (K) | 214 | 208 | 199 | 193 |
| | $E_{FM\text{-}a} - E_{FM\text{-}b}$ per sheet | 0.0286 | 0.0271 | 0.0252 | 0.0227 |
| | $E_{FM\text{-}c} - E_{FM\text{-}b}$ per sheet | 0.1958 | 0.1991 | 0.2018 | 0.2042 |

**Table s1: Calculated structural and magnetic properties of CrSBr under hydrostatic pressure.** For FM couplings, the Heisenberg magnetic exchange constants $J < 0$, while for AFM couplings, $J > 0$.

| Empirical formula | CrSBr | $CrSCl_{0.11}Br_{0.89}$ | $CrSCl_{0.27}Br_{0.73}$ | $CrSCl_{0.41}Br_{0.59}$ | $CrSCl_{0.57}Br_{0.43}$ | $CrSCl_{0.67}Br_{0.33}$ |
|---|---|---|---|---|---|---|
| Formula weight (g/mol) | 163.97 | 159.08 | 152.19 | 145.96 | 138.85 | 134.41 |
| Crystal system | Orthorhombic | Orthorhombic | Orthorhombic | Orthorhombic | Orthorhombic | Orthorhombic |
| Space group | *Pmmn* | *Pmmn* | *Pmmn* | *Pmmn* | *Pmmn* | *Pmmn* |
| Unit cell dimensions (Å,°) | a = 3.5077(4), α = 90<br>b = 4.7631(5), β = 90<br>c = 7.9591(10), γ = 90 | a = 3.4804(3), α = 90<br>b = 4.7513(5), β = 90<br>c = 7.8749(8), γ = 90 | a = 3.4756(4), α = 90<br>b = 4.7602(5), β = 90<br>c = 7.8267(10), γ = 90 | a = 3.4561(3), α = 90°<br>b = 4.7633(4), β = 90°<br>c = 7.7217(8), γ = 90° | a = 3.4418(3), α = 90<br>b = 4.7645(5), β = 90<br>c = 7.6373(8), γ = 90 | a = 3.4313(4), α = 90<br>b = 4.7667(5), β = 90<br>c = 7.5703(8), γ = 90 |
| Volume ($Å^3$) | 132.98(3) | 130.22(2) | 129.49(3) | 127.12(2) | 125.24(2) | 123.82(2) |
| Density (calculated, $g/cm^3$) | 4.095 | 4.057 | 3.903 | 3.813 | 3.682 | 3.605 |
| Independent reflections | 244 [$R_{int}$ = 0.0410] | 238 [$R_{int}$ = 0.0457] | 236 [$R_{int}$ = 0.0306] | 233 [$R_{int}$ = 0.0396] | 231 [$R_{int}$ = 0.0330] | 230 [$R_{int}$ = 0.0172] |
| Completeness to θ = 25.242° (%) | 99.4 | 99.4 | 98.7 | 99.3 | 98.7 | 99.3 |
| Goodness-of-fit | 1.051 | 1.115 | 1.017 | 1.047 | 1.021 | 1.092 |
| Final R indices [I > 2σ(I)] | $R_{obs}$ = 0.0374, $wR_{obs}$ = 0.0760 | $R_{obs}$ = 0.0401, $wR_{obs}$ = 0.0832 | $R_{obs}$ = 0.0423, $wR_{obs}$ = 0.0979 | $R_{obs}$ = 0.0480, $wR_{obs}$ = 0.1112 | $R_{obs}$ = 0.0441, $wR_{obs}$ = 0.1083 | $R_{obs}$ = 0.0210, $wR_{obs}$ = 0.0512 |
| R indices [all data] | $R_{all}$ = 0.0467, $wR_{all}$ = 0.0866 | $R_{all}$ = 0.0493, $wR_{all}$ = 0.0928 | $R_{all}$ = 0.0517, $wR_{all}$ = 0.1055 | $R_{all}$ = 0.0546, $wR_{all}$ = 0.1169 | $R_{all}$ = 0.0509, $wR_{all}$ = 0.1137 | $R_{all}$ = 0.0262, $wR_{all}$ = 0.0569 |
| Largest diff. peak and hole ($e·Å^{-3}$) | 1.498 and -1.114 | 1.416 and -1.138 | 1.523 and -1.009 | 1.399 and -1.334 | 1.206 and -1.230 | 0.583 and -0.429 |

**Table s2: Crystal data and structure refinement for $CrSCl_xBr_{1-x}$ (*x* = 0 - 0.67).** R = $\Sigma||F_o|-|F_c|| / \Sigma|F_o|$, wR = $\{\Sigma[w(|F_o|^2 - |F_c|^2)^2] / \Sigma[w(|F_o|^4)]\}11^{1/2}$ and $w=1/[\sigma^2(Fo^2)+(P)^2 + P]$ where $P=(Fo^2+2Fc^2)/3$.

| | | $CrSBr_{1-x}Cl_x$ Batches | | | | | |
|---|---|---|---|---|---|---|---|
| | | Cl-00 | Cl-11 | Cl-27 | Cl-41 | Cl-57 | Cl-67 |
| Elemental Composition | Cr | 1.00 | 1.00 | 1.00 | 1.00 | 1.00 | 1.00 |
| | S | 0.96 ± 0.03 | 0.99 ± 0.03 | 0.95 ± 0.03 | 0.92 ± 0.05 | 0.90 ± 0.02 | 0.87 ± 0.04 |
| | Br | 1.05 ± 0.03 | 0.93 ± 0.02 | 0.80 ± 0.02 | 0.59 ± 0.02 | 0.44 ± 0.02 | 0.34 ± 0.02 |
| | Cl | 0.00 | 0.13 ± 0.01 | 0.24 ± 0.01 | 0.42 ± 0.02 | 0.53 ± 0.02 | 0.68 ± 0.06 |
| | Cl/(Cl+Br) | 0.00 | 0.12 ± 0.01 | 0.23 ± 0.01 | 0.41 ± 0.01 | 0.55 ± 0.01 | 0.66 ± 0.01 |

**Table s3: Extracted chemical compositions for $CrSCl_xBr_{1-x}$ ($x$ = 0 - 0.67).** Relative concentration of Cr (blue), S (yellow), Br (red), and Cl (green) normalized to the measured Cr concentration for each $CrSCl_xBr_{1-x}$ batch. The last row is the concentration of Cl relative to the total halide concentration. Error bars represent the standard deviation. All chemical concentrations were extracted from SEM/EDX data.

| | Material (Theory/Exp) | CrSBr (Calculated) | CrSBr (Experiment) | $CrSBr_{0.33}Cl_{0.67}$ (Experiment) | CrSCl (Calculated) |
|---|---|---|---|---|---|
| Structural Properties | a (Å) | 3.516 | 3.508 | 3.431 | 3.422 |
| | b (Å) | 4.718 | 4.763 | 4.767 | 4.738 |
| | c (Å) | 8.031 | 7.959 | 7.570 | 7.506 |
| | V (Å$^3$) | 132.909 | 132.977 | 123.820 | 121.713 |
| | Cr-$S_1$ (Å) | 2.393 | 2.414 | 2.413 | 2.402 |
| | Cr-$S_2$ (Å) | 2.380 | 2.398 | 2.382 | 2.371 |
| | Cr-Br/Cl (Å) | 2.513 | 2.494 | 2.413 | 2.369 |
| | vdW gap (Å) | 2.416 | 2.384 | 2.145 | 2.191 |
| | $\Theta_2$ (°) | 96.505 | 96.38 | 96.26 | 96.592 |
| | $\Theta_{1A}$ (°) | 88.801 | 89.36 | 90.65 | 92.511 |
| | $\Theta_{1B}$ (°) | 95.210 | 94.02 | 92.12 | 92.374 |
| | $\Theta_3$ (°) | 160.657 | 161.25 | 161.91 | 160.912 |
| | $J_1$ Cr-Cr (Å) | 3.516 | 3.508 | 3.431 | 3.422 |
| | $J_2$ Cr-Cr (Å) | 3.561 | 3.586 | 3.572 | 3.564 |
| | $J_3$ Cr-Cr (Å) | 4.718 | 4.763 | 4.767 | 4.738 |
| Magnetic Properties | $E_{FM}$-$E_{AFM}$ (meV) | 0.2 | | | -0.6 |
| | $E_{FM\text{-}a} – E_{FM\text{-}b}$ per surface | 0.0286 | | | 0.0086 |
| | $E_{FM\text{-}c} – E_{FM\text{-}b}$ per surface | 0.1958 | | | 0.0804 |
| | $J_1$ (meV) | -3.165 | | | -0.624 |
| | $J_2$ (meV) | -6.495 | | | -6.739 |
| | $J_3$ (meV) | -5.702 | | | -5.488 |
| | $J_{IL}$ (meV) | 0.044 | | | -0.134 |
| | $T_C$ (K) | 214 | | | 180 |

**Table s4: Calculated structure and magnetic parameters for relaxed CrSBr and CrSCl structures along with experimental structure parameters for CrSBr and $CrSBr_{0.33}Cl_{0.67}$.**